\documentclass[designs,review,accept,pdftex,moreauthors]{Definitions/mdpi}
\usepackage{booktabs}
\usepackage{pifont}
\usepackage{multicol}
\usepackage{xcolor}
\usepackage{soulutf8}
\usepackage{float}
\firstpage{1}
\pubvolume{7}
\issuenum{4}
\articlenumber{100}
\pubyear{2023}
\copyrightyear{2023}
\externaleditor{\textls[-15]{Academic Editor: Roberto Gabbrielli}}
\datereceived{4 June 2023}
\daterevised{1 July 2023} 
\dateaccepted{13 July 2023}
\datepublished{10 August 2023 }
\hreflink{https://doi.org/10.3390/\linebreak designs7040100} 

\Title{Safety in Traffic Management Systems: A~Comprehensive~Survey}

\TitleCitation{Safety in Traffic Management Systems: A Comprehensive Survey}

\Author{Wenlu Du \href{https://orcid.org/0000-0003-4925-0468}{\orcidicon}, Ankan Dash \href{https://orcid.org/0000-0002-1151-3632}{\orcidicon}, Jing Li \href{https://orcid.org/0000-0002-6865-7290}{\orcidicon}, Hua Wei \href{https://orcid.org/0000-0002-3735-1635}{\orcidicon} and Guiling Wang *}

\AuthorNames{Wenlu Du, Ankan Dash, Jing Li, Hua Wei and Guiling Wang}

\AuthorCitation{Du, W.; Dash, A.; Li, J.; Wei, H.; Wang, G.}

\address[1]{Ying Wu College of Computing, New Jersey Institute of Technology, Newark, NJ 07102, USA; wd48@njit.edu~(W.D.); ad892@njit.edu (A.D.); jingli@njit.edu (J.L.); hua.wei@asu.edu (H.W.)}

\corres{\hangafter=1 \hangindent=1.05em \hspace{-0.82em}Correspondence: gwang@njit.edu}

\abstract{Traffic management systems play a vital role in ensuring safe and efficient transportation on roads. However, the use of advanced technologies in traffic management systems has introduced new safety challenges. Therefore, it is important to ensure the safety of these systems to prevent accidents and minimize their impact on road users. In this survey, we provide a comprehensive review of the literature on safety in traffic management systems. Specifically, we discuss the different safety issues that arise in traffic management systems, the current state of research on safety in these systems, and the techniques and methods proposed to ensure the safety of these systems. We also identify the limitations of the existing research and suggest future research directions.}


\begin{document}


\section{Introduction}
As addressed by the U.S. Department of Transportation Strategic Plan FY 2022--2026 (\url{https://www.transportation.gov/dot-strategic-plan}) (accessed on 14 March 2023),  making the transportation system safer for all people is still a top strategic goal. About 95\% of transportation fatalities in the USA occur on the country's streets, roads, and highways, and the number of deaths is increasing. Traffic safety is of paramount importance, particularly in the era of emerging technologies like automated vehicles and connected vehicles~\cite{alanazi2023systematic}. As these technologies continue to evolve and become more prevalent on the roads, the potential for safer transportation increases significantly. Automated vehicles have the potential to minimize human error, which is responsible for the majority of traffic accidents. With their advanced sensors and algorithms, they can detect and respond to potential hazards more swiftly and effectively than human drivers. Similarly, connected vehicles enable real-time communication between vehicles and infrastructure, allowing for enhanced awareness and coordination on the road. This connectivity facilitates the exchange of critical information, such as traffic conditions, weather updates, and road hazards, thereby enabling drivers to make informed decisions and avoid potential dangers. By embracing and prioritizing traffic safety in conjunction with these advanced technologies, we can strive towards a future with reduced accidents, injuries, and fatalities on the roadways, ultimately creating a safer and more efficient transportation system for all.

In the past five years, researchers have made significant efforts in the field of traffic safety~\cite{Chia2022,nascimento2019systematic,muhammad2020deep}. Some researchers, particularly those in civil engineering, have focused on statistical analysis to identify and prioritize countermeasures. By analyzing historical datasets and records, they aim to understand the cause-and-effect relationships and develop effective strategies. For instance, they examine contributory factors such as adverse weather conditions that increase the risk of accidents~\cite{mehra2020reviewnet,tobin2021effects,rahman2021assessing}. This analysis helps in devising control plans, including driver warnings, to reduce crash rates during extreme weather events. On the other hand, interdisciplinary researchers aim to provide accurate risk information by utilizing machine learning and deep learning models~\cite{muhammad2020deep}. They work towards developing real-time crash risk prediction systems. However, when it comes to operational aspects, less attention has been given to explaining the impact of variables and more emphasis has been placed on improving prediction accuracy. Techniques like deep neural networks, generative models, reinforcement learning, ensemble methods like XGBoost, and computer vision-based algorithms have gained popularity in this regard. This survey provides an overview of the advancements in these two directions, summarizing the state-of-the-art research in the field.

Furthermore, we have categorized the recent literature based on the specific areas of analysis or control. Put simply, some researchers concentrate on enhancing the overall safety of an entire traffic network or a specific region~\cite{trirat2021df,mantouka2022deep}, such as downtown New York City. Others address crash-related issues occurring on highway segments~\cite{Huang2020}, on ramp/off ramp sections~\cite{das2021saint}, weaving areas~\cite{zhao2021understand}, and curved segments~\cite{ma2020modeling}. Additionally, efforts have been made to improve safety at intersections~\cite{ghoul2021real}. As automated vehicles, connected vehicles, and connected and autonomous vehicle (CAV) technology continue to emerge, along with advanced features like automatic emergency braking (AEB) with pedestrian detection, adaptive cruise control (ACC) systems, and advanced driver assistance systems (ADAS), studies have focused on the vehicle side as well~\cite{nascimento2019systematic}. For instance, researchers evaluate the real-time risk of collisions in scenarios involving car following~\cite{zhao2022personalized} or platooning~\cite{han2022distributed}. In this survey, we also provide an overview of existing research in these aspects.

Finally, we conclude by addressing the present challenges and limitations, aiming to provide a clear understanding of areas that can be improved in the future. Through our comprehensive literature review, we observed that certain limitations are prevalent and remain unresolved to this day. One such challenge is the imbalanced data problem~\cite{Zheng2021}, which significantly complicates predictive tasks due to the limited representation of crash data within the dataset. Many researchers highlight the difficulty in collecting labeled accident data in real-world scenarios~\cite{Huang2020}. Another common issue is the lack of generalizability to real-world conditions~\cite{wang2023srl}, as some proposed models demonstrate satisfactory performance only in simulated environments, with limited evidence of successful deployment in real-world settings. It is essential to recognize and address these challenges in order to advance the field of traffic safety and improve the applicability of the research findings in practical contexts.

It is important to note the divergence between the research conducted in the field of civil engineering and interdisciplinary research, particularly in computer science. Civil engineering researchers often employ statistical analysis and sensitivity analysis to explore the correlation between variables and their impact on safety. They frequently utilize real-world datasets and conduct field tests to obtain empirical evidence. Conversely, interdisciplinary researchers tend to prioritize the design of models using simulated environments, which may or may not translate effectively into practical applications. However, there is a growing trend towards integrating domain-specific rules with popular neural network models to leverage the strengths of both approaches~\cite{du2023safelight}. This collaborative approach aims to bridge the gap and capitalize on the benefits offered by combining domain knowledge with the capabilities of neural networks.

The survey paper makes the following contributions:

\begin{itemize}
\item A thorough examination of the literature published within the last five years is conducted, allowing for an accurate depiction of the prevailing research trends during this period.

\item The literature collection exclusively focuses on top-tier venues, ensuring that the selected works are highly representative of both the domain field and computer science field. This provides valuable insights for researchers interested in traffic safety~applications.

\item We categorize the works into two distinct categories of analysis and control and provide corresponding summaries that outline the research objectives and limitations. This categorization offers inspiration and guidance for future researchers in the field.
\end{itemize}

\section{Review Method}
In this section, we outline the process of collecting our review papers. We provide details regarding the number of papers reviewed, their respective sources, and the relevant statistics. For instance, we highlight the distribution of papers between the computer science and transportation fields, shedding light on the representation from ea\mbox{ch disc}ipline.

\begin{figure}[H] \vspace{-13pt} \hspace{-16pt}
\includegraphics[width=.93\textwidth]{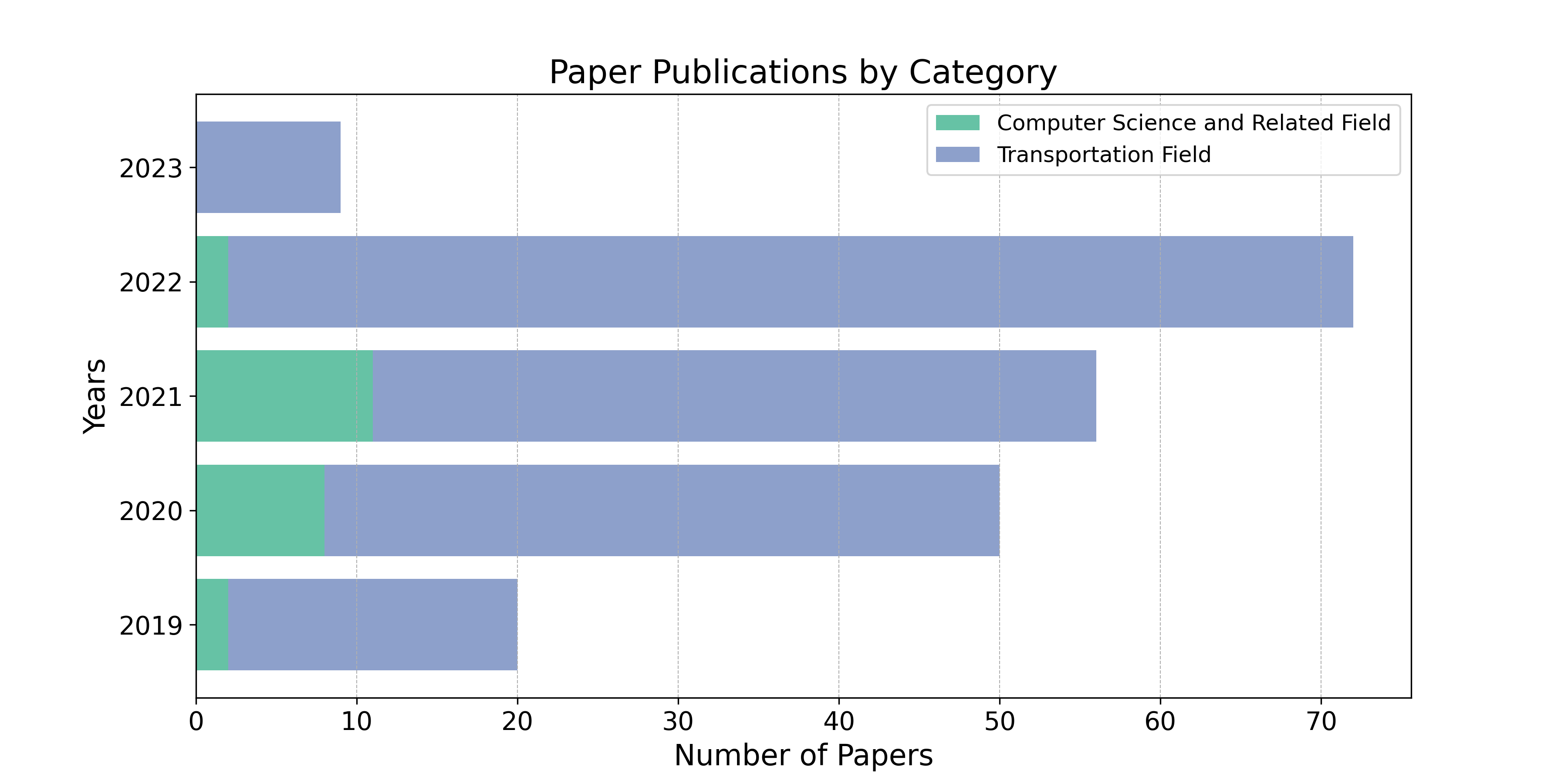} \vspace{-3pt}
\caption{Collected paper publications in different fields within the past five years.}
\label{fig:distribution_by_field}
\end{figure}

To initiate our paper collection, we conducted keyword searches for terms such as road safety, accident prevention, accident avoidance, crash risk, and traffic accident across top-tier venues in both the traditional transportation field and computer science and related disciplines. The survey covers a span of five years, specifically from 1 January 2019 to 1~June~2023. Figure~\ref{fig:distribution_by_field} presents a graphical representation of the publication distribution during this five-year period, distinguishing between publications in the transportation field and those in the computer science field. While computer science encompasses diverse research areas, we discovered several renowned venues where computer scientists contribute their work, applying proposed models to the transportation domain and demonstrating their practicality. Although the number of publications in computer science is relatively small compared to that in the transportation field, there is an evident upward trend in publications over the years, indicating the pressing need to enhance safety measures. This upward trajectory suggests that the rise in publications will likely continue in the future.

\begin{figure}[H]
\vspace{-6pt}
\includegraphics[width=.92\textwidth]{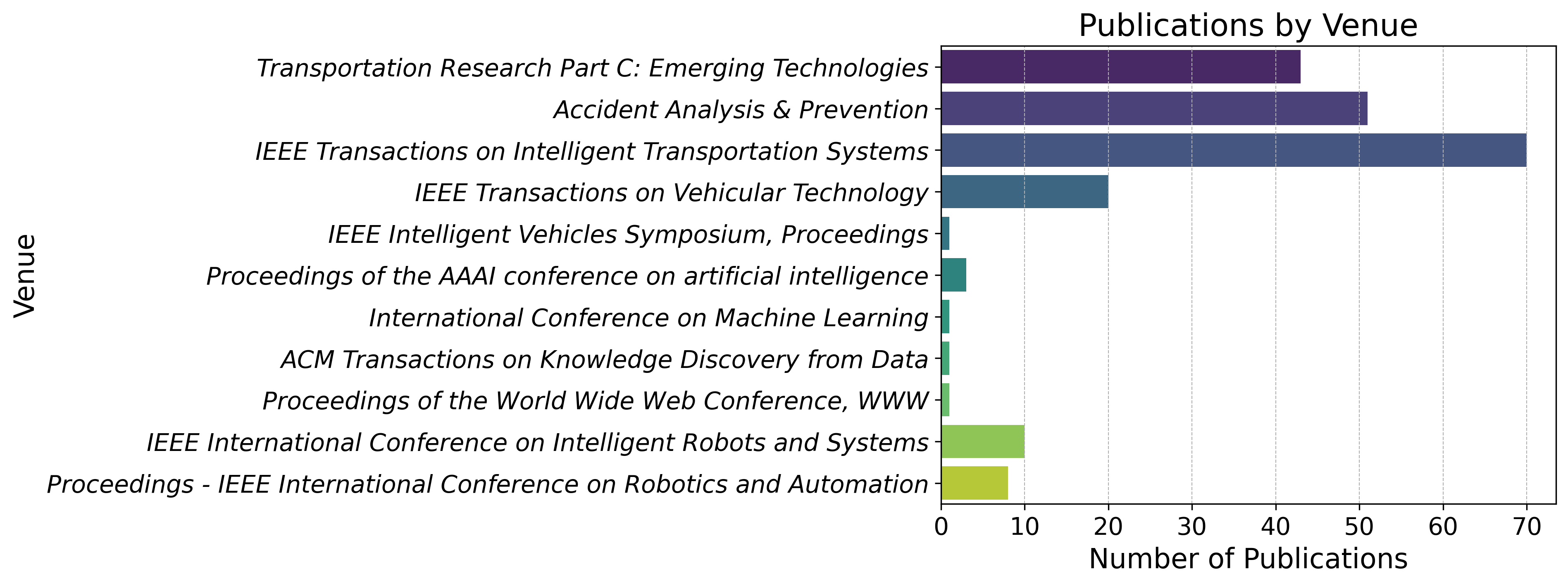} \vspace{-3pt}
\caption{Collected paper publications by publication venues within the past five years.}
\label{fig:distribution_by_venue}
\end{figure}

Figure~\ref{fig:distribution_by_venue} displays a comprehensive list of the top-tier venues utilized in this survey, along with the corresponding publication counts for each venue. In the transportation field, we observed that IEEE Transactions on Intelligent Transportation Systems accumulated the highest number of publications, suggesting its prominence among researchers for disseminating their work. 
Additionally, conferences in robotic engineering also contributed several publications, with a primary focus on autonomous driving and related techniques.

\section{Surveying the Literature: An In-Depth Exploration}
Based on the analysis of the surveyed papers, we categorized them into four distinct sections as shown in Tables~\ref{tab:summary_table} and~\ref{tab:summary_table_2}. Firstly, we highlight the recurring problems or topics that frequently appeared in the publications. Secondly, we provide a summary of the areas where safety improvements are emphasized, such as intersections or freeways. Thirdly, we examine the specific targets that researchers focused on in their efforts to enhance safety, such as connected vehicles, and we compile a comprehensive list of the trending techniques discussed in the surveyed papers.


\begin{table}[H]
\small

\caption{Keyword summary.}
\label{tab:summary_table}
\newcolumntype{C}{>{\centering\arraybackslash}X}
\begin{tabularx}{\textwidth}{CC}
\noalign{\hrule height 1pt}
\rowcolor{gray!50}
\multicolumn{2}{c}{\textbf{Topics}} \\
\hline
\begin{itemize}
\item Identification of Dangerous Vehicles
\item Accident Forecasting
\item Identification of Crash Risk
\item Crash Risk Assessment
\item Real-time Proactive Road Safety
\item Forward Collision Avoidance
\item Rear-end Collision Avoidance
\item Secondary Crash Likelihood Prediction \vspace{-9pt}
\end{itemize}&
\begin{itemize}
\item Cyclist Crash Rates Assessment
\item Rare Event Modeling
\item Inter-vehicle Crash Risk Analysis
\item Identification of High-risk Locations
\item Trajectory Predictions
\item Trajectory Collision Avoidance
\item Predictive Platoon Control
\item Pedestrian Occupancy Forecasting \vspace{-9pt}
\end{itemize}
\\
\midrule
\begin{itemize}
\item Spatial--temporal Correlations
\item Minute-Level
\item Driver Braking Behavior
\item Driver's Evasive Behavior
\item Heavy-truck Risk
\item Moving Vehicle Groups
\item School-aged Children
\item Evacuation
\item Old Drivers
\item Social Vulnerability
\item Driving Impairments and Distractions
\item Pedestrian Crash Risk Analysis
\item Automatic Emergency Braking Systems
\item Precipitation
\item Surrogate Safety Metrics
\item Adaptive Traffic Signal Control \vspace{-9pt}
\end{itemize} &
\begin{itemize}
\item Signal-vehicle Coupled Control
\item Car Following
\item Take-over Performance
\item Left-turn at Signalized Intersections
\item Safety-aware Adaptive Cruise Control
\item Adaptive Merging Control
\item Lane Keeping System
\item In-vehicle Warning
\item Context-aware
\item Multitask
\item Human Driver Imitation
\item Dashcam Videos
\item Driver Drowsiness Monitoring
\item Hazy Weather Conditions
\item On-ramp Merging Control
\item Preferences of Aggressiveness \vspace{-9pt}
\end{itemize}

\\
\bottomrule
\end{tabularx}
\end{table}\vspace{-3pt}

Upon careful observation, we identified crash risk prediction as the most extensively addressed topic among the surveyed papers. It occupied a significant portion of the literature reviewed. Furthermore, we noticed a growing trend of focusing on specific conditions or scenarios, such as heavy-truck risk, school-aged children, evacuation, extreme weather, and more. These papers aimed to address safety issues within these specific situations and propose measures for improvement. With the advent of advanced technologies, safety concerns require reassessment and reevaluation. Some works delved into the safety implications of emerging technologies, such as analyzing take-over performance or examining the impact of automatic emergency braking systems on overall safety. Additionally, certain high-risk areas that frequently experience accidents have garnered attention from researchers, leading to focused investigations on topics like on-ramp merging control. Moreover, new surrogate safety metrics have emerged as highly-discussed subjects within the literature, further reflecting the shifting landscape of traffic safety research.

It is important to acknowledge the disparity between the research conducted in the transportation domain and the research pursued by computer scientists. The traditional domain approaches primarily focused on analyzing the contributory factors leading to crashes, whereas computer science researchers were inclined towards designing more effective models for risk prediction and safety planning. In the subsequent sections, we adhere to this logical distinction by first delving into the analysis of the contributory factors and subsequently introducing various control methods.


\begin{table}[H]
\caption{Keyword summary by different perspectives.}
\label{tab:summary_table_2}
\newcolumntype{C}{>{\centering\arraybackslash}X}
\begin{tabularx}{\textwidth}{CC}
\noalign{\hrule height 1pt}
\rowcolor{gray!50}
\multicolumn{2}{c}{\textbf{Investigated Locations}} \\
\hline
\begin{itemize}
\item Freeway Segments
\item Horizontal Curvature
\item Expressways
\item Intersection  \vspace{-9pt}
\end{itemize}&
\begin{itemize}
\item Roadway Segments
\item Urban Arterials
\item Type A Weaving Segments
\item Ring Roads  \vspace{-9pt}
\end{itemize}
\\
\hline
\rowcolor{gray!50}
\multicolumn{2}{c}{\textbf{Considered Entities} }\\
\hline
\begin{itemize}
\item Connected Vehicles
\item Autonomous Vehicles
\item Cycling  \vspace{-9pt}
\end{itemize} &
\begin{itemize}
\item Motorists
\item Pedestrians  \vspace{-9pt}

\end{itemize}
\\
\hline
\rowcolor{gray!50}
\multicolumn{2}{c}{\textbf{Techniques}} \\
\hline
\begin{itemize}
\item Bayesian Network
\item Deep Reinforcement Learning
\item Reinforcement Learning Tree
\item Inverse Reinforcement Learning
\item Computer Vision
\item Matched Case Control
\item Propensity Score
\item SHapley Additive ExPlanation
\item Gradient Boosting \vspace{-9pt}
\end{itemize}&
\begin{itemize}
\item LSTM-CNN
\item Transfer Learning
\item Attention Network
\item Support Vector Machines
\item Stacked Autoencoder
\item Gated Recurrent Unit
\item Monte Carlo Tree Search
\item Imitation Learning \vspace{-9pt}
\end{itemize}
\\
\hline
\rowcolor{gray!50}
\multicolumn{2}{c}{\textbf{Data}} \\
\hline
\begin{itemize}
\item Naturalistic Driving Data
\item Simulated Data
\item Driving Simulator Platform \vspace{-9pt}
\end{itemize}&
\begin{itemize}
\item SHRP2 NDS Data
\item Event-based Data \vspace{-9pt}
\end{itemize}
\\
\bottomrule
\end{tabularx}
\end{table}

\subsection{Ongoing Funded Research Projects}
{In addition to the major scientific conferences and jounals, we also investigate active funded research projects on safety including NCHRP, FHWA, and NHTSA,  aiming to summarize the latest trend in practice.}


\textbf{NCHRP Rsearch Projects.} {A U.S. research program addressing transportation challenges, administered by TRB under the National Academies, NCHRP funds projects on various topics, including highway safety, involving experts from academia, industry, and government to enhance transportation safety. NCHRP projects aim to enhance traffic safety and develop strategies for pedestrians, bicyclists, and road infrastructure. They cover areas such as traffic safety culture, pedestrian safety, highway--rail grade crossings, rural highways, alternative intersections, motorist behavior, and leveraging AI and big data. The research focuses on improving safety, reducing crashes, and providing decision-making tools for transportation departments. Specifically, NCHRP 17-96 aims to develop a prioritized research roadmap for Traffic Safety Culture (TSC) to improve traffic safety by changing values and attitudes and strategically applying TSC strategies in collaboration with the 4Es. NCHRP 17-97 investigates the causes of nighttime pedestrian crashes, evaluates the existing and emerging strategies for improving pedestrian nighttime safety, proposes effective mitigation strategies, and develops guidance for their implementation.
NCHRP 17-99 develops a framework and tools for assessing the safety effectiveness of treatments and technologies at highway--rail grade crossings, aiding decision making to reduce incidents and improve safety. NCHRP 17-92 develops a predictive methodology for estimating the crash frequency and severity on rural two-lane two-way highways, incorporating speed measures.
NCHRP 17-109 develops Crash Modification Factors (CMFs) for Automated Traffic Signal Performance Measures (ATSPM) signal timing, quantifying safety benefits and crash reductions for all modes and conflict types. NCHRP 17-108 develops quantitative crash prediction methodologies, including Safety Performance Functions (SPFs) and Crash Modification Factors (CMFs), for alternative intersection designs (DLT, MUT, and RCUT) to quantify their safety benefits. NCHRP 17-106 quantifies the effects of centerline and shoulder rumble strips on bicyclists' safety and works to understand motorists' behavior, informing design policies and developing a guide for rumble strip applications. NCHRP 17-100 leverages AI, machine learning, and big data to provide data-driven analysis tools and prioritize investments for safer roads, focusing on pedestrians, cyclists, and new-mobility users. }

\textbf{FHWA Research Projects.} {The FHWA is a U.S. government agency that manages and improves the country's highways to make sure they are safe, efficient, and accessible. They work on projects related to road infrastructure, traffic management, and transportation planning, playing a crucial role in maintaining and enhancing the transportation network for people and goods. FHWA-PROJ-19-0014 aims to develop an Artificial Realistic Data (ARD) generator for evaluating safety analysis methods. FHWA-PROJ-20-0030 links databases to develop speed-related Crash Modification Factors (CMFs) for safety analysis. FHWA-PROJ-21-0069 uses AI models to predict traffic conditions and manage highways proactively. FHWA-PROJ-20-0054 creates a safety assessment tool for interchange designs. FHWA-PROJ-19-0089 focuses on human factors in automated vehicles. FHWA-PROJ-19-0085 evaluates intersection designs for pedestrian and bicyclist safety. FHWA-PROJ-19-0026 collects data and evaluates safety improvements for mini-roundabouts, wrong-way driving, and bicycle intersections. FHWA-PROJ-20-0002 studies the safety of pedestrian crossing signs with LEDs.}

\textbf{NHTSA Research Projects.} {The NHTSA, a U.S. federal agency under the Department of Transportation, actively promotes highway safety, sets vehicle standards, and reduces traffic injuries. It facilitates the ESV conference, a platform for sharing research and initiatives on vehicle safety, with papers published in the \textit{Traffic Injury Prevention Journal}. After reviewing the recent publications, we summarized the following studies:}

A study~\cite{atwood2023female} investigated the impact of sex on fatality rates in car crashes, finding that newer vehicles and advanced safety features have reduced fatality risks for female occupants compared to males. Another study~\cite{devane2023response} evaluated occupant models with active muscles and showed their ability to accurately predict occupant responses in crash simulations. An investigation~\cite{bolte2023analysis} focused on elderly individuals in near-side impact crashes revealed the need for further analysis in establishing injury thresholds. A study~\cite{schwarz2023multi} on drowsy-driving detection models incorporated multiple data sources and achieved good accuracy in predicting drowsiness. A study~\cite{kullgren2023effects} evaluated the crash reductions achieved in cars equipped with automatic emergency braking (AEB) systems with pedestrian and bicyclist detection. The analysis showed an overall reduction in the crash risk, with AEB systems reducing the pedestrian crash risk by 18\% and the bicyclist crash risk by 23\% during daylight and twilight conditions. However, no significant reductions were observed in darkness. Another method~\cite{breitlauch2023novel}  was developed to accurately and efficiently simulate vehicle collisions, providing collision severity parameters for injury mitigation assessment. Regulations are being developed for the safe introduction of automated driving systems, and a data-driven scenario-based assessment method was proposed~\cite{de2023certain} to estimate their safety risk.

{Through our investigation, we observed a trend towards utilizing advanced technologies, such as active muscles, AEB systems, and data-driven models, to enhance safety in various aspects of car crashes. Sex-specific analysis and understanding the impact of sex on fatality rates have gained attention. Accurate prediction, detection, and assessment of risks are crucial for enhancing safety measures. Ongoing efforts focus on developing technologies and methods for simulating and assessing collision severity, aiming to enhance injury mitigation capabilities.}

\subsection{Geographical Distribution of the Study Area}
\begin{figure}[H]
\vspace{4pt}
\includegraphics[width=.95\textwidth]{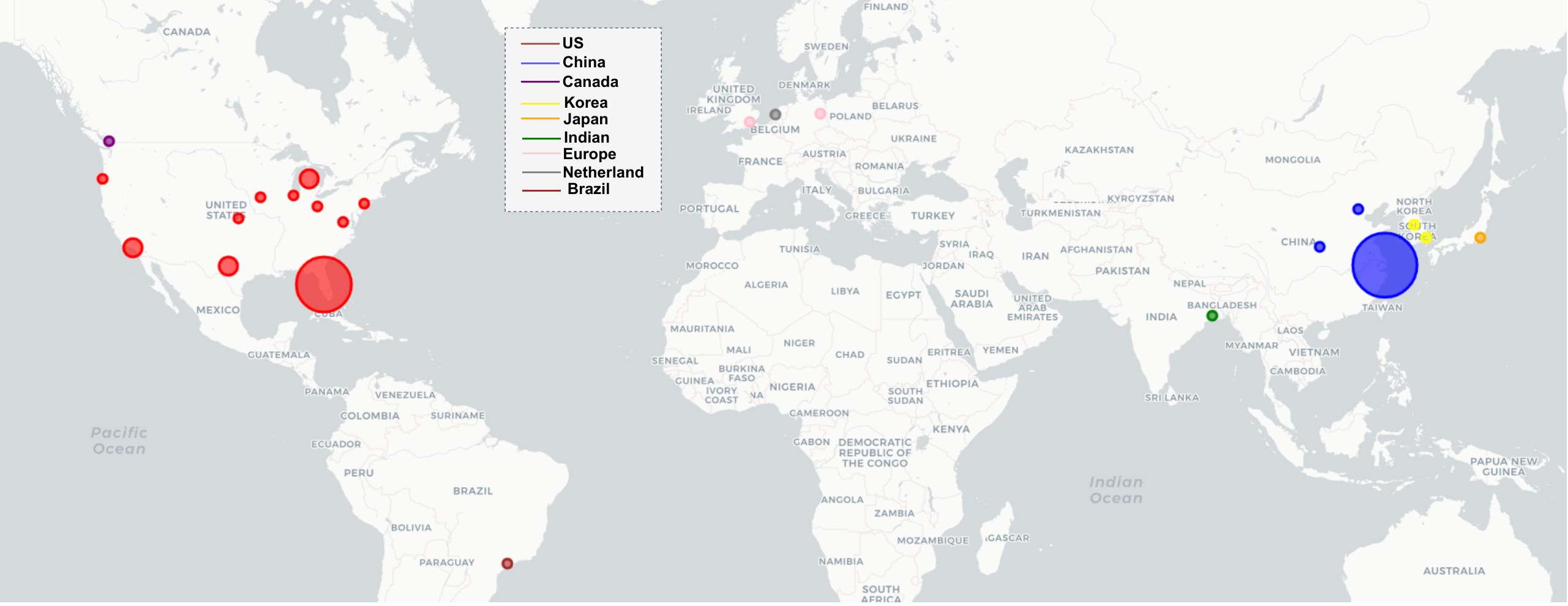}
\caption{Spatial distribution of study region: Varied colors depict diverse countries, and greater circle size signifies more extensive research in that location.}
\label{fig:geo_distribution}
\end{figure}

{We examined the research locations highlighted in the recent literature, specifically the sites where their experiments were conducted, as depicted in Figure}~\ref{fig:geo_distribution}. {Our analysis revealed that Florida, USA, and Shanghai, China, emerged as two commonly chose\mbox{n lo}cations.}

\section{Analysis}\label{S4}
The analysis of safety in traffic management systems involves evaluating and assessing the safety aspects of various components and processes within a transportation system. It aims to identify the potential hazards, assess the risks, and implement measures to mitigate those risks, ultimately ensuring the safety of road users and minimizing the occurrence of accidents. {In addition to risk analysis, researchers also strive to analyze injuries with the goal of minimizing their occurrence and severity to the lowest possible level.} The analysis of safety in traffic management systems is a multidisciplinary field that combines expertise from transportation engineering, data analysis, human factors, and policy making to ensure safer road environments and reduce the likelihood and severity of accidents {and injuries}.

\subsection{Method}
We summarize the methods used for the analysis of traffic safety.

\textbf{Matched-pair Analysis.} Matched-pair analysis, also known as paired analysis or paired comparison, is a statistical method used to compare two related sets of data or observations. It is particularly useful when studying situations where it is difficult to establish a direct cause-and-effect relationship between variables or when dealing with data that exhibit a high degree of variability. In matched-pair analysis, each observation in one group or condition is paired or matched with a corresponding observation in the other group or condition. The pairing is conducted based on similarities or relevant characteristics between the observations, such as age, sex, or some other relevant factor. The pairing ensures that each pair of observations is as similar as possible, except for the variable being investigated. By pairing observations, it helps to control for individual differences or confounding variables that could affect the outcome being measured. This analysis method increases the precision and reduces the potential biases associated with unpaired comparisons. Matched-pair analysis was applied in \cite{tobin2021effects} to analyze the relative crash risk during various types of precipitation (rain, snow, sleet, and freezing rain).

\textbf{Mutual Information Theory.} Mutual information theory is a concept in information theory that measures the amount of information that is shared or transmitted between two random variables. It quantifies the degree of dependence or association between the variables and provides a measure of the reduction in uncertainty about one variable given knowledge of the other variable. Entropy is a fundamental concept in information theory that characterizes the uncertainty or randomness of a random variable. It measures the average amount of information needed to specify the outcome of a random variable. Higher entropy indicates higher uncertainty. In addition, mutual information measures the amount of information that two random variables share. It quantifies the reduction in uncertainty about one variable by knowing the value of the other variable. Mathematically, it is the difference between the entropy of the individual variables and the joint entropy of the two variables. If the mutual information is high, it indicates a strong relationship between the variables, suggesting that knowledge of one variable provides substantial information about the other variable. Overall, mutual information theory has proven to be a valuable tool in various disciplines that deal with data analysis and information processing. Using mutual information theory, one study~\cite{baikejuli2022study} quantified the interactions between various risk factors, considering multifactor scenarios.

\textbf{Matched Case Control.} The matched case-control approach can be applied to analyze crash occurrences during special scenarios such as evacuations~\cite{rahman2021assessing,yang2022developing}. This approach allows for a thorough investigation of the potential risk factors or exposures that contribute to crashes in a special scenario while controlling for the confounding variables. For example, the authors in~\cite{rahman2021assessing}, discussed a study focused on understanding the factors contributing to increased crash occurrences during evacuations, particularly in the context of hurricanes. The researchers adopted a matched case-control approach and analyzed the traffic data collected shortly before each crash. They considered data from upstream and downstream detectors surrounding the crash location. The study included three different conditions: regular periods, evacuation periods, and a combination of both. Following is a general outline of how the matched case-control approach can be used in the analysis of crash occurrences during evacuation:
\begin{itemize}
\item Case Selection: Identify cases, which are crash incidents that occurred during evacuation events. These could include traffic accidents, collisions, or any other crash-related incidents that occurred during the evacuation process.
\item Control Selection: Select controls, which are non-crash events during the same evacuation scenario. Controls should be chosen to be comparable to cases in terms of the relevant characteristics, such as location, time, weather conditions, and traffic volume. The objective is to create pairs of cases and controls that are similar in terms of these match\mbox{ed cr}iteria.
\item Data Collection: Gather data on both cases and controls. This includes information about the evacuation scenario, road conditions, traffic management measures, driver behavior, vehicle characteristics, and any other relevant variables that may influence crash occurrences during evacuations. The data collection process can involve crash reports, eyewitness accounts, interviews, video footage, or other available sources.
\item Matching Criteria: Determine the matching criteria to create pairs of cases and controls. This could involve factors such as location, time of day, weather conditions, road type, or any other factors specific to the evacuation scenario that may contribute to crash occurrences.
\item Statistical Analysis: Perform statistical analysis to compare the exposure or risk factors between cases and controls within each matched pair. Common statistical techniques used in matched case-control studies include conditional logistic regression, which takes into account the matching and provides adjusted estimates of the association between the exposure variables and cra\mbox{sh oc}currences.
\item Interpretation: Interpret the results to identify the significant risk factors or exposures associated with crash occurrences during evacuations. The analysis should account for confounding variables and assess the strength of the associations between the identified factors and the likelihood of crashes during evacuations.
\end{itemize}

It is important to note that the success and accuracy of the matched case-control analysis rely on the availability and quality of data related to both the cases and controls. Thorough data collection and careful consideration of matching criteria are crucial to ensure valid and reliable results.

\textbf{Structural Equation Modeling (SEM).} Structural Equation Modeling (SEM) is a statistical modeling technique used to analyze complex relationships among observed and latent (unobserved) variables. It allows researchers to test and estimate the relationships between variables, examine causal relationships, and assess the overall fit of the model to the data. The researchers first specify a theoretical model that represents the hypothesized relationship between the observed variables and the latent variables. The model is typically represented as a set of equations that describe the relationships between the variables. Then, the SEM distinguishes between the observed variables and the latent variables. The relationship between variables is often depicted using a path diagram indicating the hypothesized direction and strength of the relationships. Overall, SEM is a versatile and powerful technique that can handle complex models with multiple variables, assess both the measurement and structural aspects of the model, and provide insights into the underlying relationships. Wu~et~al.~\cite{wu2021exploring} employed a sequential modeling framework using structural equation modeling (SEM) to examine the combined effects on freeway rear-end~crashes.

\textbf{Logistic Regression.} Logistic regression is a statistical modeling technique commonly used to analyze the risk factors associated with road safety. It allows researchers to understand the relationship between various risk factors and the likelihood of a specific outcome, such as road accidents or crash occurrences.

Following are the general steps of logistic regression in the context of analyzing the risk factors for road safety:
\begin{itemize}
\item Outcome Variable: First, define the outcome variable, which is typically a binary variable indicating whether an event of interest has occurred or not. For road safety analysis, the outcome variable could be a binary indicator representing whether a road accident occurred (1) or did not occur (0) for each observation or case.

\item Risk Factors: Identify the hypothesized risk factors or independent variables that are linked to the outcome variable (e.g., road conditions, driver characteristics, vehicle type, weather conditions, etc.). These risk factors can be categorical (e.g., sex and road type) or continuous (e.g., vehicle speed and age).

\item Data Collection: Gather data on the outcome variable and risk factors for each observation or case. This can involve collecting information from accident reports, police records, surveys, or any other relevant sources.

\item Model Estimation: Fit a logistic regression model to the data to estimate the relationship between the risk factors and the outcome variable. Logistic regression estimates the probability of the outcome (e.g., road accident occurrence) based on the values of the risk factors. It models the log odds or logit of the probability as a linear combination of the risk factors, using a logistic function to map the linear combination to the probability scale.

\item Interpretation of Coefficients: Estimate the coefficients for each risk factor, along with their standard errors and significance levels. These coefficients represent the log-odds ratio, indicating the direction and magnitude of the association between each risk factor and the likelihood of the outcome occurring. A positive coefficient suggests an increased likelihood of the outcome, while a negative coefficient suggests a decreased likelihood, with significance indicating the strength of the association.

\item Model Evaluation: Assess the goodness of fit of the logistic regression model and evaluate its predictive performance. Various statistical measures, such as the Hosmer--Lemeshow test, likelihood ratio test, or AIC/BIC values, can be used to evaluate the model's fit to the data.

\item Conclusion and Inference: Based on the logistic regression results, draw conclusions about the significance and impact of the risk factors on road safety. Identify the risk factors that have a statistically significant association with the outcome variable and determine their relative importance in explaining the occurrence of road accidents.
\end{itemize}

Logistic regression is a powerful tool for identifying and quantifying the relationship between risk factors and road safety outcomes. It allows researchers to understand the factors that contribute to road accidents, inform policy decisions, and develop targeted interventions to improve road safety.

The authors in~\cite{rahman2021assessing} adopted conditional logistic regression to analyze the effects of the evacuation itself on the  crash risk. 
Cicchino~\cite{cicchino2022effects} employed logistic regression to investigate the effects of automatic emergency braking with pedestrian detection on real-world pedestrian crashes. Arvin~et~al.~\cite{arvin2020driving} applied it to investigate the relationship between the duration of distractions and critical events like crashes or near-crashes. Ma~et~al.~\cite{ma2020modeling} used it to analyze the crash risk associated with highway horizontal curves. {Olszewski~et~al.}~\cite{olszewski2019investigating} {modeled accident fatality risk using binary logistic regression.} {The authors of}~\cite{hua2023injury} {employed random parameter logistic regression models with varying means and variances to assess the significant parameters affecting injury severity in reverse sideswipe collisions during the day and night over a period of nine years.} {The authors in}~\cite{wang2021risk} {examined the risk factors for severe injuries among different e-bike rider groups using a combined classification tree and logistic regression model.} The authors in~\cite{Zheng2021} employed a Bayesian logistic regression approach to identify optimal crash precursors for varying freeway section types. Bayesian logistic regression is a statistical modeling technique that combines logistic regression with Bayesian inference. It provides a framework for estimating the parameters of a logistic regression model while incorporating prior knowledge or beliefs about the parameters. Bayesian inference allows for the quantification of uncertainty and the updating of beliefs based on the observed data. Notably, aside from prior distribution and posterior distribution, Bayesian logistic regression often employs Markov Chain Monte Carlo (MCMC) sampling, such as Gibbs sampling or the Metropolis--Hastings algorithm, to obtain samples from the posterior distribution. These sampling techniques generate a large number of parameter values based on the prior, likelihood, and observed data, allowing for inference and estimation of the parameters. Bayesian logistic regression can handle small sample sizes or complex models more effectively compared to classical frequentist methods. However, Bayesian analysis typically requires more computational resources and may be more complex to implement compared to traditional logistic regression.

\textbf{Negative Binomial Regression.} Negative binomial regression is a statistical method used to analyze count data with overdispersion, which occurs when the observed variance is greater than the mean. It is a generalized linear regression model that is particularly suited for modeling count outcomes, such as the number of events or occurrences. The estimation of negative binomial regression is typically conducted using maximum likelihood estimation. The model provides estimates of the regression coefficients, which indicate the direction and magnitude of the relationship between each independent variable and the count outcome. Additionally, the model provides information on the dispersion parameter, which indicates the degree of overdispersion in the data. The authors in~\cite{zhang2021crash} utilized negative binomial regression models, using these indicators to anticipate the risk level of horizontal curve segments. Negative binomial regression was also used in~\cite{branion2020cyclist} to examine the impact of risk factors independent of exposure when analyzing the risk of cycling crashes. {The study of}~\cite{abdel2021crash} {utilized the generalized linear model with negative binomial distributions to effectively handle the dispersion present in the crash data.}

\textbf{ANOVA (Analysis of Variance).} {It is a statistical technique used to compare the means of two or more groups to determine whether there are any significant differences between them. It allows for the examination of variation within groups as well as between groups. ANOVA tests the null hypothesis that the means of all groups are equal, and if the observed differences between the groups are larger than what would be expected by chance, the null hypothesis is rejected. ANOVA provides valuable insights into the significance of group differences and is widely used in various fields, including psychology, biology, and the social sciences. The study}~\cite{wang2019road} {used ANOVA to examine accident severity but highlighted the limitations due to incomplete or unclear data in the national census and statistics yearbooks.}

\textbf{Association Rule Mining.} {Association Rule Mining (ARM) is a data mining technique that aims to discover interesting relationships or patterns within a dataset. It focuses on identifying associations or correlations between different items or variables in large datasets. The ARM works by analyzing transactions or records to find frequent itemsets, which are sets of items that often appear together. From these frequent itemsets, association rules are generated, which describe relationships between items based on their co-occurrence. These rules consist of an antecedent (if) and a consequent (then) and can be used to uncover valuable insights, make predictions, or support decision making in transportation fields. The authors of}~\cite{jiang2020analysis} {proposed an ARM-based framework with objective parameter optimization and factor extraction methods. Geographic Information System (GIS) was used for spatial analysis. The framework was applied to motorcycle accidents in Victoria, Australia, identifying the critical factors and presenting hot spots on GIS maps. The proposed framework improved the ARM performance and provided practical applications for policymakers in decision making and the severity analysis of various traffic accidents.}

\textbf{Autoencoder.} An autoencoder is an unsupervised neural network architecture that aims to learn efficient representations of input data by reconstructing them from a compressed latent space. It includes an encoder network that converts the input data into a lower-dimensional representation, as well as a decoder network that reconstructs the input based on the encoded representation. Both the encoder and decoder are trained jointly to minimize the disparity between the initial input and the reconstructed output. By doing so, autoencoders learn to capture the most salient features of the data. Zhao~et~al.~\cite{zhao2021understand} employed an autoencoder to extract the spatiotemporal features of traffic data.

\textbf{Propensity Score Weighting Approach.} The propensity score weighting approach is a statistical method used to estimate causal effects in observational studies. It addresses the issue of confounding variables by creating a weighted sample that equalizes the distribution of covariates across treatment groups, mimicking a randomized controlled trial (RCT) design. The propensity score is estimated using a logistic regression model. It summarizes the covariant information into a single value for each observation and provides  a way to create a pseudo-randomization by balancing the covariate distribution between different groups. Once the propensity scores are estimated, each observation is assigned a weight based on its propensity score. The weight reflects the inverse of the probability of the receiving control condition. One paper~\cite{lu2020evaluating} explored the causal impact of cellphone distraction on traffic accidents through the utilization of propensity score weighting methods. In this study, propensity score weighting was employed to calculate the causal odds ratio (OR) of cellphone usage across various event-populations. These populations included the overall population's average treatment effect (ATE), the treated population's average treatment effect (ATT), and the overlapping population's average treatment effect (ATO). By utilizing propensity score methods, the study achieved enhanced balance in the baseline characteristics. The findings indicate that propensity score approaches effectively address potential confounding effects arising from imbalanced driver and environmental attributes within the data.

\textbf{SHapley Additive ExPlanation (SHAP).} The SHapley Additive ExPlanation (SHAP) is a method used to explain the predictions of machine learning models. It provides an interpretation of how each input feature contributes to the model's output prediction. The SHAP is based on the concept of cooperative game theory and assigns values to each feature based on their contribution to the prediction in a fair and consistent manner. The SHapley value is calculated for each input feature, indicating its contribution to the model's prediction. The SHAP calculates feature importance by considering all possible coalitions of features and measuring their contributions to the prediction. For each coalition, the SHapley value is calculated by averaging the marginal contributions of the features. This process accounts for the interactions and dependencies between features. Additionally,  the SHAP ensures fairness and consistency by employing certain properties of the SHapley value, such as symmetry, linearity, and null player. Overall, the SHAP can be applied to various machine learning models, including black-box models like neural networks and ensemble methods to explain and interpret complex predictions. Wen~et~al.~\cite{wen2021quantifying} adopted this technique to quantify and explain risk factors on roadway segment crashes accross different crash types. Hu~et~al.~\cite{hu2020efficient} used the SHAP to measure feature importance.

\textbf{T-test.} A t-test is a statistical analysis method employed to ascertain whether there exists a notable distinction between the means of two groups or conditions. It assesses whether the difference observed in sample means is likely to be a true difference in the population or simply due to random variation. Son~et~al.~\cite{son2020effects} utilized a t-test in a connected vehicle setting to assess the efficacy of an in-vehicle advanced warning information service for mitigating secondary crashes.

\textbf{Mann--Whitney Test.} The Mann–Whitney test, also known as the Mann–Whitney U test (i.e., the Wilcoxon rank-sum test), is a nonparametric statistical test used to compare the distributions of two independent samples. It is used when the data do not meet the assumptions of normality required for parametric tests such as the t-test. The Mann–Whitney test compares the ranks of the observations between the two groups rather than the actual values. It determines whether the two samples are drawn from the same population or whether there is a significant difference between them. Following is a step-by-step overview of how the Mann–Whitney test works:
\begin{enumerate}
\item Null hypothesis (H0): The distributions of the two samples are equal. Alternative hypothesis (Ha): The distributions of the two samples are not equal.
\item Combine the data from both groups and rank them in ascending order. Assign ranks to each observation, with the lowest value assigned a rank of 1, the next lowest value assigned a rank of 2, and so on.
\item Calculate the sum of ranks (U) for each group. U1 represents the sum of ranks for one group, and U2 represents the sum of ranks for the other group.
\item Calculate the test statistic U, which is the smaller of U1 and U2. The test statistic is used to determine the \emph{p}-value.
\item Determine the critical value or \emph{p}-value associated with the test statistic. The critical value or \emph{p}-value is obtained from a reference table or statistical software.
\item Compare the calculated test statistic with the critical value or \emph{p}-value. If the calculated test statistic is less than the critical value or if the \emph{p}-value is less than the predetermined significance level (e.g., $\alpha$ = 0.05), the null hypothesis is rejected, indicating that the two groups have a significant difference. If the calculated test statistic is greater than the critical value or if the \emph{p}-value is greater than the significance level, the null hypothesis is not rejected, suggesting that there is no significant difference between the t\mbox{wo gro}ups.
\end{enumerate}

Torok~et~al.~\cite{torok2022automated} applied the Mann--Whitney test to analyze the crash risk involving automated vehicles.

\subsection{Research Outcomes}
We thoroughly investigated and analyzed the collected papers, and we provide a comprehensive summary of the research outcomes regarding the analysis approaches employed in traffic management systems' safety. Specifically, we condense the findings pertaining to the various factors contributing to crashes or accidents. The results are categorized based on the examined locations or entities, as illustrated in Table~\ref{tab:summary_table_2}.

\textbf{Analysis of Roadway Safety.}
Tobin~et~al.~\cite{tobin2021effects} highlighted that the relative crash risk is significantly higher during periods of precipitation compared to non-precipitation periods. In a similar vein, Wen~et~al.~\cite{wen2021quantifying} demonstrated that the importance of risk factors varies across different crash types. For rear-end (RE) crashes, the speed limit emerged as a more crucial risk factor than lane width, right/left shoulder width, and median width. Conversely, for run-off-road (ROR) crashes, the opposite relationship was observed. Additionally, the study revealed that narrow lanes (8 ft to 11 ft) elevated the risk for all types of crashes, while a lane width of 12 ft or more in road segments with five or six lanes in both directions combined may aid in mitigating the risk of all types of crashes.

\textbf{Analysis of Freeway Safety.}
Several research findings have contributed to our understanding of factors influencing crash occurrence and risk mitigation. Rahman~et~al.~\cite{rahman2021assessing} identified a high variation in speed at a downstream station and high traffic volume at an upstream station as factors increasing the likelihood of crash occurrence. Zheng~et~al.~\cite{Zheng2021} further supported this by demonstrating that the downstream average speed was the best crash precursor variable across different segment types. The effectiveness of warning information systems in preventing secondary crash risks was shown by Son~et~al.~\cite{son2020effects}, with Jang~et~al.~\cite{nguyen2020turn} reporting a significant reduction in the crash potential through the provision of warning information. Wu~et~al.~\cite{wu2021exploring} emphasized the importance of speed oscillation patterns and driver response time, as prolonged response time due to distractions can increase the crash risk. Zhang~et~al.~\cite{zhang2021crash} highlighted the effectiveness of in-vehicle data, such as lateral and longitudinal stability indicators, in assessing the road crash risk. Ma~et~al.~\cite{ma2020modeling} demonstrated the significance of including elevation features to address the confounding impact of vertical curves along H-curves. Finally, Ding~et~al.~\cite{ding2019structural} stressed the importance of incorporating visual perception, including speed risk perception and distance risk perception, and suggested the potential application of line markings. These findings collectively contribute to a deeper understanding of crash factors and offer insights into potential strategies for mitigating crash risks on the freeway segments. {From the perspective of injury analysis, the authors of}~\cite{hua2023injury} {investigated the influence of the time of day on the injury severity and identified the factors that contributed to the severity. Positive associations were found between factors like male driver, reckless behavior, and adverse roadway conditions, while factors such as older driver, residential area, and wet road surface were found to mitigate the severity. The study also highlighted the temporal instability and time-of-day fluctuations, emphasizing the need for segmentation. Countermeasures like centerline rumble strips and intelligent vehicle technologies were recommended to mitigate the injury severity.}

\textbf{Analysis of Intersection Safety.}
Several research findings have shed light on various aspects related to crash risk and intersection safety. Kwon~et~al.~\cite{kwon2020examination} emphasized the significance of intersection characteristics, such as the proportional area of sky and roadway, in influencing the perceived crash risk among school-aged children. Mitra~et~al.~\cite{Mitra2020} identified multiple factors that significantly influenced both the frequency and severity of crashes, including blocked carriageways, approach traffic volume, traffic configuration, type of minor road, presence of protected right turning phase, tram stops, all-red time, visibility of road markings, and non-motorized traffic. Essa~et~al.~\cite{Essa2019} highlighted the temporal aspect of intersection safety, revealing that the highest frequency of conflicts occurred at the start of the green time, while the greatest severity was observed at the beginning of the red time. Furthermore, Zafian~et~al.~\cite{Zafian2021} emphasized the importance of considering alternative data sources and collection methods to address the literature gap in improving safety for older drivers. These findings collectively contribute to a deeper understanding of the factors influencing crash risk at intersections and provide insights for developing effective safety strategies.

\textbf{Analysis of Vehicle, Pedestrian, and Cyclist Safety.}
Research findings in the field of road safety have provided insights into various factors affecting the crash risk for different road users. For vehicle-related factors, Baikejuli~et~al.~\cite{baikejuli2022study} highlighted the contribution of multifactor interaction, such as environmental and vehicular factors, in increasing the crash probability for heavy-truck fatal crashes. Mattas~et~al.~\cite{mattas2020fuzzy} emphasized the potential of the real-time evaluation of rear-end collision risk using fuzzy surrogate safety measures. Arvin~et~al.~\cite{arvin2020driving} identified alcohol and drug impairment, as well as distractions associated with activities like cellphone use, as significant contributors to crash/near-crash events. Lu~et~al.~\cite{lu2020evaluating} confirmed the increased crash risk associated with cellphone use, with visual--manual tasks posing a higher risk than talking on a cellphone.

Regarding driver-related factors, Rowe~et~al.~\cite{rowe2022newly} identified risky driving style, skill deficiencies, and driving confidence as factors derived from the early driving development questionnaire. Lin~et~al.~\cite{lin2020understanding} found differences between lower-crash-risk (LCR) and higher-crash-risk (HCR) drivers in terms of the brake reaction times and hazard perception, highlighting the role of task engagement in driving performance.

For cyclists, Branion~et~al.~\cite{branion2020cyclist} identified the factors associated with higher crash risks, including less frequent cycling, male sex, negative perceptions of cycling in the neighborhood, and residing in areas with high building density. Torok~et~al.~\cite{torok2022automated} revealed that crashes were more severe when the autopilot mode was turned off in automated vehicles.

Lastly, Ko~et~al.~\cite{ko2021multi} demonstrated the safety benefits of connected vehicles, showing a significant reduction in the crash potential index (CPI) in Connected Vehicle--Connected Vehicle cases compared to Regular Vehicle--Regular Vehicle cases. These findings collectively contribute to our understanding of crash risk factors and offer insights for developing effective strategies to enhance road safety for different road users.

\subsection{Comparative Analysis of the Literature: Commonalities and Variations}
We performed keyword analysis on studies published within the last five years and categorized them according to the publication venue.

\begin{figure}[H]
\vspace{-3pt}
\includegraphics[width=.95\textwidth]{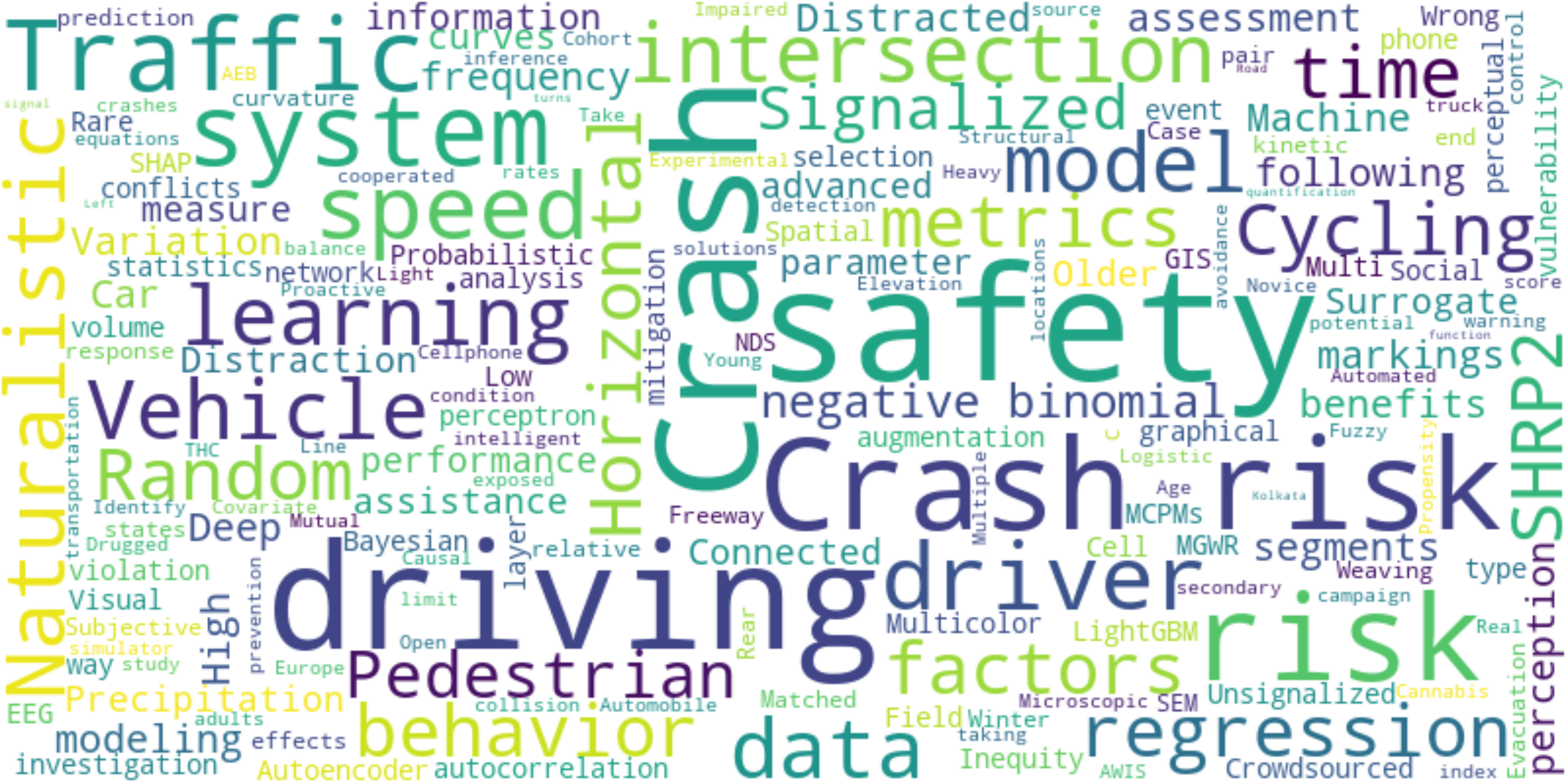}
\caption{Keyword cloud for studies published in Accident Analysis and Prevention.}
\label{fig:wordcloud_1}
\end{figure}

(1) The resulting keyword cloud for studies published in Accident Analysis and Prevention is depicted in Figure~\ref{fig:wordcloud_1}. 
Out of the 45 publications examined, the Analysis category comprised 32 studies.

Furthermore, we provide a summary of the titles, highlighting the common themes found in these studies:
\begin{itemize}
\item Crash risk assessment: Several titles focused on assessing crash risk using different methodologies, such as data-driven Bayesian networks, quantification of risk factors, and identification of high-risk segments.

\item Impact of driving behavior: Multiple titles explored the effects of various driving behaviors on crash risk, including evasive behavior, distraction, impaired driving, and subjective risk perception.

\item Utilization of data: Many titles utilized real-world data or simulation platforms to predict collision cases, examine crash risks, and identify high-risk locations.

\item Specific contexts: Some titles investigated crash risk in specific contexts, such as urban cycling, freeway segments with horizontal curvature, and signalized intersections.
\end{itemize}

The variations among the literature included:
\begin{itemize}
\item Methods and techniques: Each title employed different methods and techniques for crash risk assessment, ranging from Bayesian networks and deep reinforcement learning to empirical observations and structural equation modeling.

\item Focus areas: The titles covered a wide range of topics within the domain of crash risk, including individual driver assessment, roadway segment crashes, intersection safety, driving behaviors, and the impact of factors like distraction and precipitation.

\item Data sources: The titles made use of diverse data sources, such as traffic violation and crash records, EEG metrics, naturalistic driving data, connected vehicle systems data, and SHRP2 NDS data.
\end{itemize}

Overall, these titles collectively highlight the importance of data-driven approaches, the role of specific risk factors and behaviors in crash occurrence, and the potential for using advanced techniques to assess and mitigate crash risks in various contexts.

\begin{figure}[H]
\vspace{4pt}
\includegraphics[width=.95\textwidth]{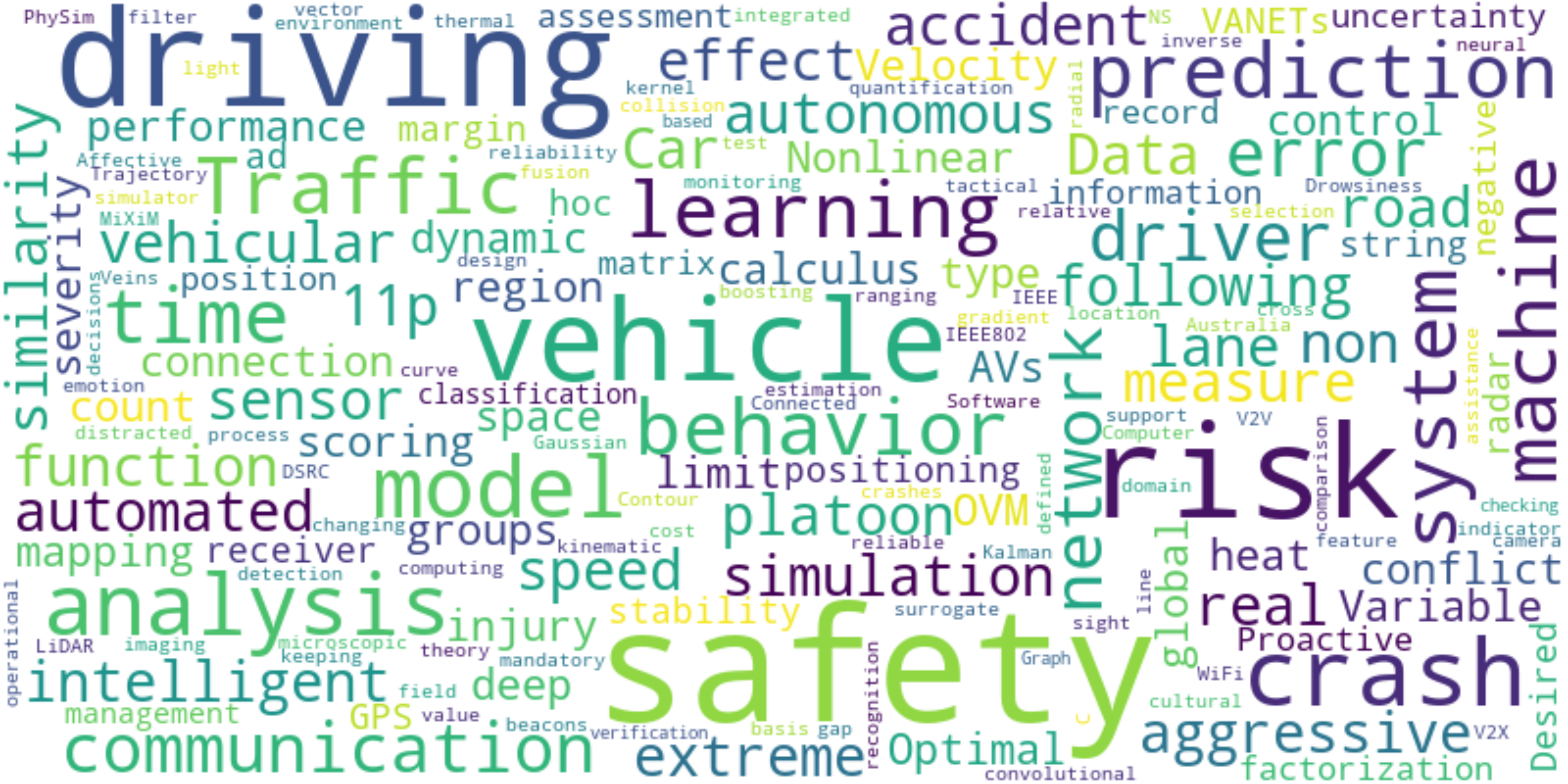}
\caption{Keyword cloud for studies published in IEEE Transactions on Intelligent Transportation Systems (ITSC).}
\label{fig:wordcloud_2}
\end{figure}

(2) The resulting keyword cloud for studies published in IEEE Transactions on Intelligent Transportation Systems is depicted in Figure~\ref{fig:wordcloud_2}.
Out of the 67 publications examined, the Analysis category comprised 23 studies.

The common themes observed in these titles included:
\begin{itemize}
\item Risk Analysis and Safety Assessment: Several studies focused on analyzing driving risks, crash avoidance, and safety evaluation in different contexts, such as variable speed limit systems, automated driving strategies, and vehicular safety applications.

\item Driving Behavior Analysis: Several studies explored driving behavior patterns, habitual driving behaviors, and driver interactions with vehicle systems, aiming to understand their impact on crash risk and safety.

\item Comparative Analysis and Evaluation: Several studies compared and evaluated different algorithms, models, or features related to safety applications, such as LiDAR-based contour estimation, driver drowsiness monitoring, and vehicular communications.
\end{itemize}

The differences among these titles lay in the specific contexts and methodologies employed. For example, the studies varied in terms of the location of analysis (e.g., urban expressways in Shanghai), the focus on specific technologies (e.g., software-defined vehicles or connected environment), and the analytical approaches used (e.g., Bayesian analysis, nonnegative matrix factorization, and behavioral anomaly detection).

We conducted a comparison of the titles published in Accident Prevention and Analysis with those from ITSC, and we found the following key findings:

There was a common focus in both lists on the assessment and quantification of crash risks. These studies explored factors such as driver behavior and roadway conditions and the impact of external events like weather and evacuation. Additionally, the analysis of driving behavior and distractions as they related to crash risk was present in both lists. This included factors like driver evasive behavior, the duration of distractions, and the influence of cellphone usage on the crash probability.

However, there were also notable differences between the two venues. The titles from ITSC placed emphasis on specific contexts such as individual driver risk assessment, wrong-way driving crashes, urban cycling, and freeway rear-end crashes. In contrast, the studies published in Accident Analysis and Prevention focused on topics such as signalized intersections, heavy-truck crash risk, and the impact of traffic signal performance.

Moreover, the titles in Accident Analysis and Prevention explicitly mentioned specific analytical techniques and methods such as Bayesian networks, deep reinforcement learning, EEG metrics, propensity score methods, and random parameter modeling. These techniques were not explicitly mentioned in ITSC.

Additionally, the studies published in Accident Analysis and Prevention included titles that examined crash risks and safety measures in specific locations like Texas, seven European cities, Kolkata, and Korean freeways. On the other hand, the titles from ITSC did not mention specific geographic locations.

Overall, while there were some common themes between the two lists, they differed in terms of the specific contexts, analytical techniques, and geographic scope.

\section{Operation or Control}
This section focuses on the proactive methods to control traffic to enhance safety. These methods aim to prevent accidents and ensure the overall safety of road users. In terms of intersection management, proactive measures involve the use of traffic signals, roundabouts, and clear signage to enhance safety. Additionally, advanced traffic signal systems that utilize sensors and algorithms have been implemented to optimize the traffic flow and minimize the likelihood of collisions. Proactive traffic management also encompasses the use of intelligent transportation systems (ITS) to efficiently manage and control traffic. This includes real-time traffic monitoring, dynamic message signs, traffic cameras, and incident management systems, which promptly detect and respond to accidents or congestion. Furthermore, researchers utilize data from traffic management systems, including historical accident data and real-time traffic data, to identify high-risk areas and predict potential accident hotspots. This valuable information can guide targeted safety interventions. Implementing a combination of these proactive safety measures can significantly enhance traffic management and reduce the risk of accidents on the roads.

\subsection{Method}
We summarize the methods used to control traffic to enhance safety.

\textbf{Convolutional Neural Networks (CNNs).} Convolutional Neural Networks (CNNs) are deep learning models specifically designed for analyzing visual data like images. They use layers of filters to extract meaningful features from the input images. These filters perform convolution operations, highlighting patterns and structures. The network then learns to recognize complex features by stacking multiple convolutional layers. Pooling layers downsample the feature maps to capture important information. Fully connected layers process the extracted features and make predictions. CNNs are trained using labeled data, optimizing their parameters to minimize prediction errors. CNNs have revolutionized computer vision tasks by automatically learning relevant features directly from images, enabling them to achieve high accuracy in tasks like image classification and object detection. In one study~\cite{Huang2020}, the authors treated the spatiotemporal traffic data as an image and included the time-of-day information as an extra node in fully connected layers to determine whether a specific traffic condition was prone to crashes or not. Hu~et~al.~\cite{hu2020efficient} showed that sensor data obtained from connected vehicles (CVs) could be treated as a one-dimensional image, and the features hidden in the vehicle kinetic and traffic data could be extracted and learned by a CNN model. Triat~et~al.~\cite{trirat2021df} applied CNN to learn a hidden representation of each district’s static environmental features. Zhao~et~al.~\cite{zhao2021understand} employed a Gated Convolutional Network (G-CNN) to identify different traffic states and their associated crash risks. G-CNN has been developed recently, as a variant of the traditional CNN architecture that incorporates gating mechanisms inspired by recurrent neural networks (RNNs). Gated CNNs aim to capture long-range dependencies within the spatial dimensions of an input image or sequence.

\textbf{LSTM.} LSTM (Long Short-Term Memory) is a type of neural network that excels at processing sequential data, like text or time series. Unlike traditional recurrent neural networks (RNNs), LSTMs can remember long-term dependencies in the data by selectively retaining or discarding information using memory cells and specialized gates. This allows them to capture patterns and make accurate predictions in tasks involving sequences. Previous research often combines the convolutional layers with the long short-term memory (LSTM) neural network in a unified deep learning framework, utilizing LSTM to extract temporal features. In one study~\cite{wang2020safedrive}, CNN-LSTM was employed to evaluate the safety risk in merging situations. Li~et~al.~\cite{li2020real} utilized an LSTM-CNN network architecture for real-time prediction of crash risk on arterials. Both the CNN and LSTM modules simultaneously receive and independently learn the data. The parallel and sequential LSTM-CNN, which were proposed in previous studies, were compared. The model took into account various features including signal timing, traffic flow characteristics, and weather conditions.

\textbf{XGBoost.} XGBoost, short for Extreme Gradient Boosting, is a powerful machine learning algorithm known for its exceptional performance in various predictive modeling tasks. It belongs to the gradient boosting family of algorithms and combines the strengths of decision trees with a boosting approach. XGBoost iteratively builds an ensemble of weak decision tree models, optimizing a specific objective function by minimizing the residuals of the previous model. It incorporates regularization techniques to prevent overfitting and employs parallel processing to accelerate training. XGBoost is highly efficient, scalable, and capable of handling large-scale datasets, making it a popular choice for tasks such as regression, classification, and ranking. One study~\cite{li2022hybrid} employed XGBoost as the model for predicting both primary and secondary crashes.

\textbf{Reinforcement Learning (RL).} Reinforcement learning is a branch of machine learning, where an agent learns to make sequential decisions in an environment to maximize a cumulative reward signal. It involves an agent interacting with an environment, taking actions, receiving feedback in the form of rewards, and adjusting its behavior over time through trial and error. Reinforcement learning is applicable to traffic management because it can optimize traffic flow, reduce congestion, and improve overall efficiency. By treating traffic management as a sequential decision-making problem, reinforcement learning algorithms can learn to control traffic signals, dynamically adjust the traffic flow, and optimize the traffic patterns based on real-time feedback. This adaptive and proactive approach can lead to more effective traffic management strategies, reduced travel times, and improved safety on the roads.

Ghoul~et~al.~\cite{ghoul2021real} selected Soft Actor--Critic (SAC) as the RL agent to optimize safety. The work~\cite{gong2020multi} applied multiobjective reinforcement learning (MORL) to simultaneously improve mobility and safety at the signalized intersection. Similarly, Du~et~al.~\cite{du2023safelight} integrated domain rules into an existing backbone RL model to enhance safety at intersections. Wang~et~al.~\cite{wang2023srl} applied RL methods coupled with safety constraints and expert strategies for the trajectory-tracking control problem. Das~et~al.~\cite{das2021saint} introduced a dual RL agent-based method to achieve an optimal tradeoff between traffic efficiency and driving safety/comfort. This was accomplished by adjusting the safety model parameters and the inter-vehicle gap according to real-time traffic data. Yang~et~al.~\cite{yang2022developing} employed a reinforcement learning tree to determine the importance of variables for real-time crash risk prediction. Mantouka~et~al.~\cite{mantouka2022deep} used RL, more specifically the DDPG agent, to personalize driving recommendations that improve driving safety while considering individual driving styles and preferences. Zhu~et~al.~\cite{zhu2020safe} employed DDPG using a reward function that combined driving features related to safety, efficiency, and comfort, referencing human driving data. Cao~et~al.~\cite{cao2022trustworthy} proposed a trustworthy improvement RL that combined RL with existing rule-based algorithms in autonomous vehicles.

\textbf{Generative Adversarial Network.} GAN stands for Generative Adversarial Network, which is a type of deep learning model consisting of two main components: a generator and a discriminator. The generator is trained to produce synthetic data samples that exhibit similarity to the real data, while the discriminator is trained to differentiate between the real and synthetic data. They are trained simultaneously in a competitive manner, with the goal of the generator producing data that can fool the discriminator.

In the context of imbalanced data in traffic management, GANs can be utilized to address the issue of limited or imbalanced crash record data~\cite{man2022wasserstein}. By training the GAN on the available crash data, the generator can learn to generate synthetic crash records that resemble the real ones. This can help augment the existing imbalanced dataset with additional samples, thereby increasing the representation of underrepresented crash scenarios. By generating synthetic data, GANs provide an effective means to balance the distribution of crash records, enabling more robust and accurate models to be developed for traffic safety analysis, risk assessment, and proactive safety interventions.

\textbf{Graph Neural Network.} Graph Neural Networks (GNNs) are a type of neural network specifically designed to handle data with graph structures. GNNs can effectively capture and model the relationships between entities represented as nodes and edges in a graph. In the context of traffic management, GNNs can be applied to various safety-related tasks. For instance, GNNs can analyze road networks and capture the complex dependencies between different road segments, traffic intersections, and their associated attributes (e.g., traffic volume and speed limits). By leveraging this information, GNNs can predict traffic congestion, identify accident-prone areas, and even optimize traffic signal timings for improved safety. GNNs enable a holistic view of the traffic system by considering the spatial relationships and interactions between road elements, leading to more accurate and context-aware safety predictions and interventions in traffic management.

In Zhou~et~al., 2020~\cite{zhou2020riskoracle}, the authors utilized a Graph Convolutional Network (GCN) for traffic accident prediction, leveraging the ability of the GCN to model non-Euclidean subregion-wise propagations and correlations. Wang~et~al.~\cite{wang2021gsnet} employed GCN to explore the spatial--temporal geographical correlations among regions based on geography.

\textbf{Transformer.} The Transformer model is a powerful neural network architecture primarily used for sequence-to-sequence tasks, such as machine translation and natural language processing. Unlike traditional recurrent neural networks (RNNs), Transformers rely on self-attention mechanisms, enabling them to capture global dependencies between input elements efficiently. The model comprises an encoder and a decoder, each consisting of multiple layers of self-attention and feed-forward neural networks. The Transformer's attention mechanism allows it to attend to relevant parts of the input sequence, facilitating parallel processing and capturing long-range dependencies effectively. In the context of proactive traffic control methods, Transformers can be applied to tasks such as traffic flow prediction and optimization. By analyzing historical traffic patterns and real-time data, Transformers can capture complex spatiotemporal dependencies, learn traffic dynamics, and make accurate predictions. The ability to model global dependencies and process large-scale data efficiently makes Transformers suitable for proactive traffic control, enabling more effective and timely interventions to manage traffic flow, mitigate congestion, and enhance overall safety on the roads. The study conducted by Trirat~et~al.~\cite{trirat2021df} incorporates a Transformer layer with an attention mechanism to emphasize the crucial time period within the temporal input features.

\subsection{Research Outcomes}

\textbf{Control Methods for Highway, Roadway, and Urban Arterials.}
\textls[-15]{In general, researchers} have explored strategies for enhancing deep learning models to attain improved performance, while simultaneously addressing the challenge of data imbalance and devising optimization techniques for it. Less complex deep models were found to achieve better performance according to Huang~et~al.~\cite{Huang2020}, while Li~et~al.~\cite{li2020real} demonstrated that their LSTM-CNN model achieved superior performance in terms of the AUC value, false alarm rate, and sensitivity compared to other models. Another study by Li~et~al.~\cite{li2022hybrid} showed that a hybrid model with fewer features achieved higher true-positive rates and a very low false alarm rate, making it suitable for real-time proactive traffic safety systems. Additionally, Zhao~et~al.~\cite{zhao2021understand} found that crash severity and types differed among various traffic states. By incorporating the heterogeneity of crash mechanisms across these states, the utilization of G-CNN demonstrated enhanced performance. Furthermore, Xie~et~al.~\cite{xie2019use} confirmed that high-risk locations identified based on connected vehicle data from a comparatively shorter duration were comparable to those identified using historical crash data. In addressing imbalanced data, Peng~et~al.~\cite{PENG2020105610} demonstrated that undersampling, the Synthetic Minority Oversampling Technique (SMOTE), and the ensemble approach (Rusboost algorithm) improved performance. Finally, Man~et~al.~\cite{man2022wasserstein} recommended the use of Wasserstein Generative Adversarial Networks (WGAN) over other oversampling methods for handling imbalanced datasets, achieving a crash prediction sensitivity of approximately 70\% with a false alarm rate of 5\%. To summarize, the literature extensively covers discussions on model design, encompassing the integration of various crash mechanisms, utilization of effective datasets, and selection of suitable network architectures. Additionally, several studies have specifically addressed the challenge of data sparseness by employing generative models.

\textbf{Control Methods for Intersections.}
Researchers have made notable contributions in optimizing safety and efficiency at signalized intersections. Ghoul~et~al.~\cite{ghoul2021real} proposed a signal-vehicle coupled control system that effectively enhanced safety with low computational intensity. Hu~et~al.~\cite{hu2020efficient} demonstrated that deep learning models, particularly the CNN model with an accuracy of 93.8\%, are recommended for predicting risk levels at intersections by combining CV data and deep learning networks. Even with low CV penetration rates, this approach showed promise in determining crash risks. Additionally, Zhang~et~al.~\cite{9130932} addressed the challenge of the real-time optimization of signal green timing and the coordination of Connected Automated Vehicle (CAV) trajectories, resulting in significant reductions in delays and stopping times of 50\% to 97\%, while eliminating collision risks. In general, there is a growing tendency to integrate signal and vehicle control systems. Moreover, the utilization of connected vehicle (CV) data has been demonstrated as a valuable approach to enhance safety measures.

\textbf{Control Methods for Road Regions.}
In recent studies, several approaches have been proposed to address different aspects of traffic risk prediction. Zhou~et~al.~\cite{zhou2020riskoracle} devised a dynamic graph neural network to capture real-time traffic variations and correlations among subregions, showcasing the efficacy of multitask learning. Wang~et~al.~\cite{wang2021gsnet} combined a spatial--temporal geographical module with Graph Convolutional Networks (GCN) and Gated Recurrent Units (GRU), showcasing improved performance. Trirat~et~al.~\cite{trirat2021df} utilized a combination of a Convolutional Neural Network (CNN), GRU, and Transformer to predict citywide traffic accident risk, with the inclusion of dangerous driving statistics proving beneficial. Furthermore, Mantouka~et~al.~\cite{mantouka2022deep} highlighted that while self-aware driving suggestions may enhance individual driving behavior, they do not necessarily improve overall traffic conditions. After examining the existing research, it is evident that the integration of a geographical module capable of capturing the spatial and temporal relationships within road regions holds significant promise. Approaches such as Graph Convolutional Networks (GCN) and neural networks with attention mechanisms have demonstrated effectiveness in this regard. However, challenges persist in dealing with the complexity arising from multiple influencing factors, multiscale dependencies, and the rarity of accident events, which require further attention and resolution in future studies.

\textbf{Control Methods for Vehicles.}
In the field of autonomous driving, there is a general preference for models with transferability, primarily due to the fact that a significant portion of the existing literature relies on simulated environments rather than real-world scenarios. Wang~et~al.~\cite{wang2023srl} achieved impressive results by successfully transferring their one-shot learning approach across simulation and realistic scenarios, showcasing a low average running time and minimal lateral error during field tests. Building on this, Wang~et~al.~\cite{wang2020safedrive} investigated the varying influences of driving patterns exhibited by surrounding vehicles in different positions on the evaluation of driving risk, highlighting the strong influence of cross positions followed by diagonal cross positions. In a related study, Kim~et~al.~\cite{kim2019crash} developed a synthetic data generator based on a driving simulator, validating the effectiveness of their proposed dangerous vehicle classifier through real-data experiments, which significantly reduced the missed detection rate compared to classifiers trained solely on real data. Furthermore, Das~et~al.~\cite{das2021saint} proposed a dual reinforcement learning (RL) agent-based approach aimed at striking an optimal balance between driving safety/comfort and traffic efficiency, achieving notable enhancements in traffic flow, driving safety, and overall comfort when compared to state-of-the-art methods. Lastly, Cao~et~al.~\cite{cao2022trustworthy} successfully combined RL with existing rule-based algorithms in autonomous vehicles, surpassing arbitrary rule-based driving policies and harnessing the advantages of learning-based methods in stochastic scenarios, while also ensuring trustworthy safety improvements derived from rule-based policies. In summary, multitask learning, one-shot transfer, hybrid models, and the integration of rule-based driving policies have emerged as dominant and effective methods in the field.

\subsection{Comparative Analysis of the Literature: Commonalities and Variations}
\begin{figure}[H]
\vspace{-3pt}
\includegraphics[width=.98\textwidth]{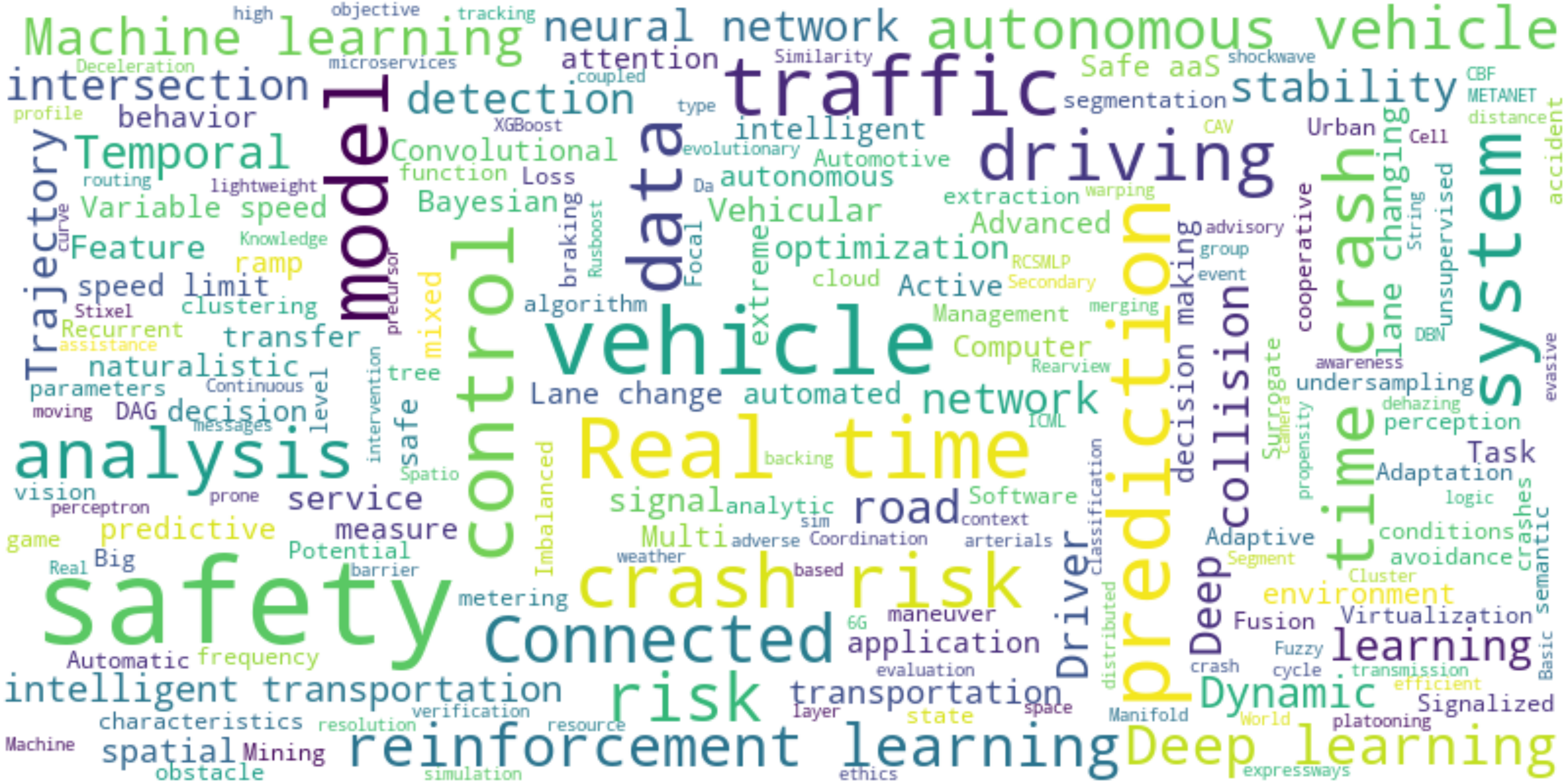}
\caption{Keyword cloud for control methods.}
\label{fig:wordcloud_3}
\end{figure}

We obtained the keywords of 74 surveyed control methods and represented them visually using a word cloud, as depicted in Figure~\ref{fig:wordcloud_3}.

As in Section \ref{S4}, we proceeded to examine the titles of these control methods. The titles encompassed various topics related to road safety, intelligent transportation systems, and crash risk prediction. Common themes observed in these titles included real-time crash risk prediction, reinforcement learning-based frameworks for road safety management, intelligent intervention systems, machine learning models for crash likelihood prediction, and the use of deep learning techniques for analyzing traffic data.

Additionally, there was a focus on specific areas of study, such as highway segments, urban expressways, intersections, and freeway bottlenecks. Different approaches and techniques were explored, including transfer learning, hybrid machine learning models, deep reinforcement learning, Bayesian networks, convolutional neural networks, and spatio--temporal graph representation learning.

These titles also highlight the importance of factors such as driver behavior, traffic flow characteristics, adaptive signal control, secondary crash prediction, and the use of connected vehicle data. Furthermore, there was an emphasis on safety measures and risk assessment in the context of autonomous vehicles, including topics like collision avoidance, personalization of car following, ethical algorithms, and the safety of cooperative driving.

Overall, these titles demonstrate a diverse range of research efforts aimed at enhancing road safety, utilizing advanced technologies, and developing intelligent systems to mitigate crash risks in various traffic scenarios.

\section{Crash Risk Prediction}
In this section, we provide an elaborate description of crash risk prediction methods as they constitute a substantial portion of the control category. This section of the paper provides a comprehensive review of the existing literature on crash risk prediction, covering various methodologies, predictive factors, etc., employed in the field.

Zheng~et~al.~\cite{ZHENG2020102683} introduced a method for forecasting the real-time probability of crashes at signalized intersections, focusing on individual signal cycles. Their approach relied on extracting traffic conflicts from informative vehicle trajectories as a foundation for crash prediction. To address the challenges posed by nonstationarity and unobserved heterogeneity in their model, they established a Bayesian hierarchical structure. The main contribution of their research lay in their novel measurement of traffic conflicts. Specifically, they employed computer vision techniques to extract traffic conflicts, quantified as modified time to collision, along with three cycle-level traffic parameters (shock wave area, traffic volume, and platoon ratio) from video data. In a subsequent study by Gu~et~al.~\cite{Gu2022-lz}, they further explored the realm of intersection safety by highlighting the advantages of connected vehicle technology. This innovation offers abundant vehicle motion data, establishing a stronger link between crash occurrence and driving behaviors. The authors also addressed the challenge of spatial dependence in crash frequency and the multitude of driving features involved in the prediction process. The novelty of their research lay in the introduction of a new artificial intelligence technique known as Geographical Random Forest (GRF). This technique effectively handled spatial heterogeneity and incorporated all the potential predictors. The researchers successfully applied the developed GRF model to predict the occurrence of rear-end crashes at intersections. Additionally, in the context of intersection safety, Lin~et~al.~\cite{9424477} devised a high accident risk prediction model through examining traffic accident data to identify risk factors specifically at intersections.

Basso~et~al.~\cite{BASSO2018202} developed accident prediction models for a section of the Autopista Central urban expressway using disaggregated data from Automatic Vehicle Identification (AVI) toll gates. The paper emphasized the use of unbalanced data and multiple repetitions for validation, which contributed to its real-time application feasibility. The conclusions highlighted the importance of vehicle type-specific variables. Cai~et~al.~\cite{CAI2020102697} presented an evaluation of real-time crash prediction models, with a particular focus on expressway data. They highlighted the persistent challenge of the extreme imbalance between crash and non-crash traffic data. The researchers expressed concerns regarding previous studies that may overlook the heterogeneity within the non-crash data due to undersampling, which is one strategy commonly used to address the imbalance issue. The noteworthy aspect of their research lay in the utilization of a deep convolutional generative adversarial network (DCGAN) model. This model effectively balanced the dataset by generating synthetic crash-related data, thus utilizing the entire non-crash data. In comparison to the minority oversampling technique (SMOTE) and random undersampling technique, the DCGAN model exhibited a superior ability to capture the characteristics of crash data, resulting in the highest prediction accuracy. Additionally, only the DCGAN-based model identified the significant impact of speed difference between upstream and downstream locations. Similarly, Peng~et~al.~\cite{PENG2020105610} addressed the issue of data imbalance and highlighted that previous studies primarily concentrated on algorithm-level solutions. In response, they proposed a three-level optimization method with a particular emphasis on the output level. The output level optimization involved the use of Youden index methods and the probability calibration method. The data level optimization incorporated the common SMOTE technique. Lastly, the algorithm level optimization explored the cost-sensitive MLP algorithm and Adaboost algorithm. On the other hand, Huang~et~al.~\cite{Huang2020} underscored the significance of crash risk prediction in mitigating secondary crashes on highways. They specifically emphasized the superior performance of deep models (i.e., CNNs with dropouts) in crash detection while also shedding light on the challenge of accurately predicting a crash risk for a traffic condition 10 min before an actual crash occurred. Notably, their approach incorporated real-time traffic data obtained from roadside radar sensors, including information on volume, speed, and sensor occupancy. Additionally, as highlighted in the study by Guo~et~al.~\cite{GUO2021106328}, previous research has predominantly concentrated on real-time crash prediction for expressways and freeways using traffic flow data, disregarding the impact of risky driving behavior. To bridge this gap, Guo~et~al.~\cite{GUO2021106328} employed in-vehicle AutoNavigator software to capture the relevant data. Li~et~al.~\cite{li2022hybrid} offered a unique viewpoint in the realm of crash risk prediction. They expressed criticism of existing studies that predominantly focused on predicting the likelihood of primary crashes leading to secondary crashes, while neglecting the likelihood of the occurrence of secondary crashes in itself. To address this limitation, they introduced a hybrid model consisting of one XGBoost model for predicting the likelihood of primary crashes and another for predicting the likelihood of secondary crashes. The proposed model aimed to forecast the risk within a short timeframe (e.g., \mbox{5--10 min}) and allowed for regular updates of the predictions on a minute-by-minute basis. A variable speed limit (VSL) system aimed at improving urban expressway safety in real time was proposed by Roy~et~al.~\cite{ROY2022106512}. The study applied reinforcement learning (RL) for VSL control. Similarly, Yang~et~al.~\cite{yang2022developing} introduced the Reinforcement Learning Tree (RLT) for real-time crash risk prediction and automatic crash detection on urban expressways. The proposed framework utilized large disaggregated datasets, and the study emphasized the significance of collecting more data and selecting relevant variables to enhance predic\mbox{tive pe}rformance.

In addition to crash risk prediction on highways, class imbalances are also present in the prediction of driving safety risks. To address this issue, Chen~et~al.~\cite{8617709} introduced a novel approach using a deep autoencoder network with L1/L2-nonnegativity constraints and cost sensitivity. This method effectively handled class imbalances and enhanced the prediction performance by determining the optimal sliding window size and automatically extracting hidden features from driving behaviors. The limitation of previous studies in analyzing driver factors and driving maneuvers, due to the absence of disaggregated driving or accident data, was highlighted by Mahajan~et~al.~\cite{9310713}. To overcome this issue, they introduced a method that considered both the likelihood and potential severity of a collision. This study focused on estimating the rear-end crash risk in specific traffic states and emphasized the significance of comprehending the evolution of crash risk under diverse traffic conditions for real-time crash prediction. Taking advantage of the progress made in deep learning technology, Li~et~al.~\cite{9858680} proposed an attention-based LSTM model for predicting lane change behavior. Their objective was to enhance both the accuracy and interpretability. Their approach involved two components: a prejudgment model utilizing a C4.5 decision tree and bagging ensemble learning and an LSTM model with an attention mechanism for multistep lane change prediction. Chen~et~al.~\cite{CHEN2020102646} presented a preemptive Lane-Change Risk Level Prediction (P-LRLP) method for estimating the crash risk during lane change maneuvers. The method used machine learning classifiers and key space-series features to predict risk levels before the maneuver was completed. An innovative resampling method (EST) and advanced classifier (LightGBM) were employed. The study recommended specific position considerations and highlighted the potential integration with ADAS and V2V communication. In a recent study, Karim~et~al.~\cite{9732278} carried out early traffic accident anticipation from dashcam videos using a dynamic spatial--temporal attention network. Specifically, a gated recurrent unit (GRU) was trained alongside attention modules to predict accident probabilities. The paper identified opportunities for integrating computer vision with other technologies and methods for safety enhancement. Formosa~et~al.~\cite{9733966}, in their research, directed their attention towards advanced driver assistant systems (ADAS) and highlighted the uncertainties associated with the deployment of connected and autonomous vehicles into heterogeneous traffic environments. They emphasized the limitations of previous studies that predominantly relied on predefined movement patterns and a single factor (time to collision) to estimate the threat levels. To address these limitations, the paper introduced the utilization of deep learning models that considered a variety of factors to estimate threat levels and predict conflicts amidst uncertainty, with a specific focus on incorporating the concept of looming. Following a similar line of inquiry, Arbabzadeh~et~al.~\cite{7945483} presented a data-driven methodology for the real-time prediction of traffic safety risk in advanced driver assistant systems (ADAS). The study utilized data from the SHRP 2 naturalistic driving study and employed elastic net regularized multinomial logistic regression to construct predictive models. Notably, driver-specific variables were incorporated into the models to enable customization.

A rear-end collision prediction method for smart cities was proposed by Wang~et~al.~\cite{9122456}. The CNN model was used for prediction using real trajectory data, while synthetic oversampling using the genetic theory of inheritance was employed to address the class imbalance. Trirat~et~al.~\cite{trirat2021df} introduced a deep gusion network for predicting traffic accident risk across urban areas, integrating hazardous driving statistics collected from in-vehicle sensors. The study examined the correlation between dangerous driving offenses and traffic accidents, revealing a strong correlation in terms of the location and time. The work by Elassad~et~al.~\cite{ELAMRANIABOUELASSAD2020102708} addressed the importance of real-time crash prediction and the development of fusion frameworks for intelligent transportation systems. They explored the use of machine learning models and fusion techniques to improve crash predictions by considering diverse data sources, including driver inputs, physiological signals, vehicle kinematics, and weather conditions. The paper highlighted the significance of addressing the class imbalance and presented the effectiveness of boosting in combination with k-NN, Naïve Bayes, Bayesian networks, and SVM with MLP as the meta-classifier. Zhou~et~al.~\cite{zhou2020riskoracle} presented a framework for accurate real-time traffic accident forecasting at minute-level granularity, which is crucial for public safety and urban management. Existing methods often face challenges due to the dynamic nature of road networks and the scarcity of accident records. Therefore, in this paper, the authors proposed RiskOracle, a novel framework that improved minute-level accident forecasting. They addressed the zero-inflated issue and sparse sensing by transforming zero-risk values in labels and introducing the DTGN (Differential Time-varying Graph Neural Network) to capture instantaneous variations and inter-subregion relationships. Additionally, the authors employed multitask and region selection schemes to identify high-risk accident subregions. In a similar vein, Wang~et~al.~\cite{wang2021gsnet} introduced GSNet, a novel model designed to address the zero-inflation issue in traffic accident risk forecasting. GSNet incorporated both geographical and semantic aspects to learn spatial--temporal correlations effectively. The model consisted of a spatial--temporal geographical module and a spatial--temporal semantic module, which captured the relevant correlations, along with the utilization of a weighted loss function. Hao~et~al.~\cite{9774974} presented an enhanced active safety prediction approach that utilized big data and a stacked autoencoder--gated recurrent unit (SAE-GRU) to predict the safety levels based on the recognition results.

Stülpnagel~et~al.~\cite{VONSTULPNAGEL2020105584} examined the relationship between objective crash risks and subjective risk perception in urban cycling. Data from a medium-sized German city, including objective crash data and subjective reports through crowdsourcing, were linked to infrastructure and traffic properties. The findings revealed a disparity between the subjective risk perception and the actual crash risk, with certain locations and situations being perceived as more or less risky than their objective risk suggested. Understanding these disparities can inform the design of safer cycling infrastructures and promote cycling as a comfortable mode of transportation.

To summarize, there have been several attempts at data-driven crash risk prediction. Some papers focused on using traditional statistical analysis and machine learning techniques \cite{Gu2022-lz, ZHENG2020102683, 7945483}, while others utilized deep learning \cite{CAI2020102697, Huang2020}, computer vision \cite{9732278}, and reinforcement learning \cite{ROY2022106512, yang2022developing} for crash risk prediction. Dealing with imbalanced data is a primary concern in crash risk prediction, and papers often addressed techniques for removing the class imbalance \cite{CAI2020102697} or developing algorithms that can handle imbalanced data \cite{PENG2020105610, Gu2022-lz}. Additionally, some works focused on incorporating driver behavior data to improve the crash risk prediction accuracy \cite{GUO2021106328, SHANGGUAN2022106500}. Moreover, identifying variables such as road width, speed limit, speed drop, lane changing, and roadside markings plays an integral role in crash risk prediction, and active research has been conducted in this area as well \cite{9424477, 9858680}.

\section{Challenges and Limitations}
In this section, we present an overview of the common challenges and limitations that researchers encountered. Furthermore, we delve into how these challenges were effectively addressed in their respective works.

\subsection{Sparseness of Data}
A prevalent issue in crash studies is the sparseness of crash data or an  imbalanced dataset, which refer to the limited availability of detailed crash records. This sparsity poses challenges in accurately analyzing crash patterns, identifying risk factors, and developing effective safety interventions. The scarcity of crash data can result from various factors, including underreporting, data collection limitations, and low-frequency crash occurrences. This limited data availability hampers the ability to capture the full spectrum of crash scenarios, leading to potential biases and inaccuracies in statistical analyses. Researchers have addressed this issue by employing various techniques such as data imputation, statistical modeling approaches, and incorporating supplementary data sources to compensate for the lack of detailed crash data. These efforts aim to enhance the robustness and reliability of crash studies, allowing for more informed decision making in traffic safety management. As addressed in~\cite{Huang2020, Zheng2021}, a thorough examination of the model's structure is necessary, particularly when working with a limited data size.

\textbf{Sampling Strategies.} The matched case-control design represents a conventional technique for undersampling. However, in \cite{li2020real}, the authors suggested that employing this conventional approach may potentially hinder the model's performance when applied to real-world data. In addition, certain valuable information pertaining to non-crash events might be overlooked or omitted during the training phase. As an optimized approach, the synthetic minority oversampling technique (SMOTE) was proposed. This method only affects the training dataset, while the testing dataset can still reflect the real-world information.  Some other variants of this method are the SVM SMOTE and \mbox{SMOTE + ENN}. SMOTE is one popular method and has been applied in several studies~\cite{li2020real,PENG2020105610,man2022wasserstein} to improve the performance when dealing with imbalanced data. Another popular sampling method is the Adaptive Synthetic Sampling (ADASYN) approach. The ADASYN focuses on the minority class by generating synthetic samples that are strategically created in regions where the class imbalance is more severe. The key difference between the ADASYN and the SMOTE lies in their synthetic sample generation process. While the SMOTE creates synthetic samples by linearly interpolating between existing minority class instances, the ADASYN uses a density distribution-based approach. The ADASYN calculates the density distribution of minority class samples and generates synthetic samples in regions where the density is lower. This adaptive nature of the ADASYN allows it to pay more attention to challenging instances and provides better handling of the class imbalance problem.

\textbf{The Ensemble Approach.} This approach helps capture important patterns and relationships even when data are limited, leading to more accurate predictions and insights about crash risk factors and contributing to improved traffic safety management. For example, gradient tree boosting has several mechanisms to handle sparse data. First, it can handle missing values by intelligently splitting the data based on the available features. Second, it can assign higher weights to the rare samples or underrepresented classes, allowing the algorithm to prioritize learning from sparse instances. This adaptive weighting helps to mitigate the impact of data sparsity on the overall model performance. The Adaptive Boosting method (i.e., AdaBoost) is another notable traditional example. AdaBoost is known for its ability to adaptively learn from data by assigning higher importance to challenging samples. It is robust against overfitting and can effectively handle imbalanced datasets. By combining weak models and focusing on difficult instances, AdaBoost can create a powerful ensemble model that achieves high predictive accuracy across a variety of machine learning tasks.

As mentioned in~\cite{man2022wasserstein}, ensemble methods are not intended to handle an imbalanced dataset. Instead they are used for improving classification performance. Thus, researchers often combine an ensemble approach with sampling methods together.

\textbf{Artificial Intelligence Models (Generative Models, Variational Autoencoder, etc.).} In recent times, artificial intelligence (AI) models have shown the capability to address the problem of data imbalance by generating synthetic data. One popular example is the utilization of variational autoencoders (VAEs) and convolutional autoencoders, as evidenced in studies by Zhao~et~al.~\cite{zhao2021understand}, Chen~et~al.~\cite{8617709}, Hao~et~al.~\cite{9774974}, and Islam~et~al.~\cite{islam2021crash}. However, VAEs may not produce samples that are as realistic as those generated by generative models such as Generative Adversarial Networks (GANs), primarily due to the use of the L2 loss function. Consequently, researchers have redirected their focus towards GANs for oversampling crash cases in more recent investigations~\cite{CAI2020102697,man2022wasserstein,basso2021deep}. As demonstrated in~\cite{man2022wasserstein}, the Wasserstein Generative Adversarial Network (WGAN) was employed and outperformed other oversampling methods in handling the imbalanced~dataset.

\subsection{Model Interpretability}
ML models face criticism for being ``black-boxes'' as they lack transparency and interpretability. This hampers their widespread adoption in safety modeling. Interpretable models allow traffic engineers and stakeholders to understand the underlying factors that influence safety outcomes, such as the identification of high-risk areas, critical contributing factors to accidents, or potential interventions to improve safety. By having interpretable models, safety methods can be effectively evaluated, validated, and fine-tuned, leading to more reliable and trustworthy decision making. SHapley Additive exPlanations (SHAP) have been presented in several studies~\cite{wen2021quantifying,hu2020efficient} in recent years to address the model interpretability issue.

\subsection{Real-World Generalizability}
Certain studies opt for conducting experiments in simulated environments rather than real-world settings, enabling the development of numerous advanced technologies. However, this choice raises concerns regarding the generalizability of the findings to real-world scenarios. This concern is particularly evident when examining adaptive traffic signal control (ATSC) methods applied at intersections. Ghoul~et~al.~\cite{ghoul2021real} devised a signal-vehicle coupled control system, utilizing PTV VISSIM simulation software to replicate the intersection environment. Some studies~\cite{9130932, das2021saint,mantouka2022deep} also employed other popular traffic simulation software such as SUMO (Simulation of Urban Mobility), an open-source microscopic coding platform capable of simulating multimodal traffic on extensive road networks. Note that the issue of sim-to-real transfer problems has been widely acknowledged as a significant concern when it comes to employing reinforcement learning methods. This poses a major challenge that practitioners utilizing RL technology must carefully consider. On the other hand, several studies have endeavored to examine traffic safety through the utilization of a driving simulator. Employing the driving simulator, an adaptive curve speed warning (ACSW) system was formulated, which provides speed warnings to drivers when navigating curved road sections based on individual driver perception and reaction times~\cite{ahmadi2019drivers}. A multiuser driving simulator was utilized to examine the pattern of rear-end collisions within vehicle platoons under foggy weather conditions and different speed limits~\cite{huang2020using}. A multiagent driving simulator (MADS) was also used to quantify traffic safety effects for each vehicle pair taking into account the interactions between vehicles in a connected vehicle environment~\cite{ko2021multi}. To validate the proposed concepts and control structures for HDV (heavy-duty vehicle) platooning, a realistic co-simulation approach was employed~\cite{Gratzer2022}. This involved utilizing high-fidelity vehicle dynamics simulation software, specifically IPG TruckMaker, to simulate each individual vehicle. MATLAB served as the simulation environment, while Simulink was utilized as the communication interface connecting the individual vehicle instances. The proposed method for intelligent navigation systems incorporated driving behavior data from the target vehicle as well as the surrounding vehicles, leading to the simulated implications~\cite{wang2020safedrive}.

\textbf{Sim-to-Real Transfer.} To address the problem of generalizability, a lightweight adapter, known as a one-shot transfer module, was proposed~\cite{Gratzer2022} and has demonstrated successful transferability between simulation and real-world scenarios.

\subsection{Limited Data Source}
The limited availability of data is a critical challenge that significantly impacts the performance in various ways. In the study conducted by Lu~et~al.~\cite{lu2020evaluating}, imbalanced driver and environmental characteristics were observed in the dataset. Huang~et~al.~\cite{Huang2020} addressed the issue of inaccurate labels in the provided data, while Li~et~al.~\cite{Huang2020} mentioned the unavailability of real-time data. Kim~et~al.~\cite{kim2022meta} stated the challenges associated with collecting labeled accident data in the real world. The heterogeneity of crash data was highlighted in the work by Baikejuli~et~al.~\cite{baikejuli2022study}, and Ghoul~et~al.~\cite{ghoul2021real} asserted that intersections with different characteristics may exhibit diverse effects. Measurement errors in vehicle trajectory data were noted by Xie~et~al.~\cite{xie2019use}. Ma~et~al.~\cite{ma2020modeling} discussed the scalability of the data, while studies by Yang~et~al.~\cite{yang2022developing}, Torok~et~al.~\cite{torok2022automated}, and Ko~et~al.~\cite{ko2021multi} identified the sample size as a major concern.

\textbf{Synthetic Dataset via Simulators.} The concept of utilizing simulators or video games to generate a vast amount of synthetic data with labels has proven to be beneficial in tackling the challenge of limited data availability. In~\cite{kim2019crash}, the authors manipulated driving agents within the virtual realm, a popular video game named Grand Theft Auto V (GTA V), to deliberately create dangerous driving scenarios that are not commonly observed in the real world. Du~et~al.~\cite{du2023safelight} also used the traffic simulator, SUMO, to deliberately create aggressive driving scenarios.

\section{Conclusions}
We conducted a systematic review of recent advancements in proactive safety methods. Our analysis encompassed the relevant literature published in top-tier venues over the past five years. We categorized the papers into analysis and control categories to provide a comprehensive understanding of the subject. Our investigation identified the trending statistical analysis methods, and we compiled the findings from these studies. Additionally, we summarized the popular control methods, noting that these approaches often work synergistically to achieve superior performance. Notably, methods such as LSTM-CNN, Reinforcement Learning, and the newly developed Transformer have been specifically designed to address safety concerns in traffic management systems. Furthermore, we highlighted the challenges encountered in the existing works, particularly regarding the scarcity of crash data and the difficulty in acquiring and labeling such data. We believe that this survey will offer valuable insights for future studies aiming to develop safety-aware traffic management systems.

\vspace{6pt}

\funding{This research was funded by Federal Highway Administration's Exploratory Advanced Research (EAR) Program under Grant Number 693JJ320C000021NAME, and by National Science Foundation (USA) under Grant Numbers CNS--1948457 and IIS-2153311.}

\dataavailability{Not applicable.}

\conflictsofinterest{The authors declare no conflict of interest. The funders had no role in the design of the study; in the collection, analyses, or interpretation of data; in the writing of the manuscript; or in the decision to publish the results.}
\newpage

\begin{adjustwidth}{-\extralength}{0cm}

\reftitle{References}

\PublishersNote{}
\end{adjustwidth}
\end{document}